\documentclass[twocolumn]{aastex63}
\usepackage[utf8]{inputenc}
\usepackage{graphicx}
\usepackage{color}

\DeclareGraphicsExtensions{.ps,.eps,.pdf,.jpg,.png}

\newcommand{\kms}{\ifmmode{\rm km\thinspace s^{-1}}\else km\thinspace s$^{-1}$\fi}
\newcommand{\rsun}{R$_\odot$}
\newcommand{\msun}{M$_\odot$}
\newcommand{\eb}{2M0056$-$08}
\newcommand{\teff}{\ensuremath{T_{\rm eff}}}
\newcommand{\vsini}{$v \sin i$}
\newcommand{\vmac}{$v_{\rm mac}$}

\begin{document}


\title{An Eclipsing Binary Comprising Two Active Red Stragglers of Identical Mass and Synchronized Rotation: A Post--Mass-Transfer System or Just Born That Way?}
\author[0000-0002-3481-9052]{Keivan G.\ Stassun}
\affiliation{Department of Physics and Astronomy, Vanderbilt University, Nashville, TN 37235, USA}
\author[0000-0002-5286-0251]{Guillermo Torres}
\affiliation{Center for Astrophysics $\vert$ Harvard \& Smithsonian, 60 Garden Street, Cambridge, MA 02138, USA}
\author[0000-0002-5365-1267]{Marina Kounkel}
\affiliation{Department of Physics and Astronomy, Vanderbilt University, Nashville, TN 37235, USA}
\author[0000-0003-2053-0749]{Benjamin M. Tofflemire}
\altaffiliation{51 Pegasi b Fellow}
\affiliation{Department of Astronomy, The University of Texas at Austin, Austin, TX 78712, USA}
\author[0000-0002-3944-8406]{Emily Leiner}
\affiliation{Department of Physics, Illinois Institute of Technology, Chicago, IL 60616, USA}
\affiliation{Center for Interdisciplinary Exploration and Research in Astrophysics (CIERA) and Department of Physics and Astronomy, Northwestern University, 1800 Sherman Ave., Evanston, IL 60201, USA}
\author[0000-0002-2457-7889]{Dax L.\ Feliz}
\affiliation{Department of Astrophysics, American Museum of Natural History, New York, NY 10024, USA}
\author[0000-0001-6977-9495]{Don M. Dixon}
\affiliation{Department of Physics and Astronomy, Vanderbilt University, Nashville, TN 37235, USA}
\author{Robert D. Mathieu}
\affiliation{Department of Astronomy, University of Wisconsin, Madison, WI 53706, USA}
\author[0000-0002-8443-0723]{Natalie Gosnell}
\affiliation{Colorado College}
\author[0000-0002-4020-3457]{Michael Gully-Santiago}
\affiliation{Department of Astronomy, The University of Texas at Austin, Austin, TX 78712, USA}

\begin{abstract}
We report the discovery of \eb\ as an equal-mass eclipsing binary (EB), comprising two red straggler stars (RSSs) with an orbital period of 33.9~d. Both stars have masses of $\approx${1.419}~\msun, identical to within {0.2}\%. Both stars appear to be in the early red-giant phase of evolution; however, they are far displaced to cooler temperatures and lower luminosities compared to standard stellar models. The broadband spectral energy distribution shows NUV excess and X-ray emission, consistent with chromospheric and coronal emission from magnetically active stars; indeed, the stars rotate more rapidly than typical red giants and they evince light curve modulations due to spots. These modulations also reveal the stars to be rotating synchronously with one another. There is evidence for excess FUV emission and long-term modulations in radial-velocities; it is not clear whether these are also attributable to magnetic activity or if they reveal a tertiary companion. Stellar evolution models modified to account for the effects of spots can reproduce the observed radii and temperatures of the RSSs. If the system possesses a white dwarf tertiary, then mass-transfer scenarios could explain the manner by which the stars came to possess such remarkably identical masses and by which they came to be sychronized. However, if the stars are presumed to have been formed as identical twins, and they managed to become tidally synchronized as they evolved toward the red giant branch, then all of the features of the system can be explained via activity effects, without requiring a complex dynamical history.
\\
\end{abstract}

\section{Introduction}
Eclipsing binary (EB) star systems have long served as empirical touchstones for theories of stellar structure and evolution. 
The best characterized EBs can produce measurements of stellar masses and radii that are precise and accurate to the percent level \citep[e.g.,][]{Torres:2010}. When compared against grids of stellar models, the predictions of stellar theory can be stringently tested, key input parameters in stellar models refined, and missing physical ingredients in the models identified and empirically constrained.
For example, EBs have been central to the development of stellar evolution models that incorporate the effects of strong surface magnetic fields on the bulk properties and internal structures of low-mass stars at young ages \citep[see, e.g.,][and references therein]{Stassun:2004,Stassun:2006,Stassun:2008,Stassun:2012,Stassun:2014,Feiden:2016,David:2019,Somers:2020,Tofflemire:2022,Stassun:2022}.

Stars in binary-star systems can also evolve along pathways that depart entirely from standard, single-star evolutionary theory, due to stellar interactions, mass transfer, and mergers. These effects are especially salient during the subgiant and giant stages of evolution, as the stars swell and their surfaces come into closer proximity. 

Sub-subgiant stars (SSGs) are an unusual class first identified in the color-magnitude diagrams of well-studied open clusters \citep[see, e.g.,][and references therein]{Mathieu:2003}. SSGs lie to the red of the main sequence and fainter than the red giant branch, a region not easily populated by standard stellar evolution pathways or by any combination of the light of two normal stars. 
More luminous objects that appear to be more like red giants, but still displaced far to the red of the normal red giant branch, are sometimes referred to as Red Straggler stars (RSSs). 
\citet{Mathieu:2003} suggested that SSGs/RSSs may be products of close stellar encounters involving binaries, and other authors have also invoked mass transfer and stellar collisions \citep[e.g.,][]{Albrow:2001,Hurley:2005}.
\citet{Leiner:2017} examined possibilities including mass transfer in a binary system, stripping of a subgiant's envelope during a dynamical encounter, and reduced luminosity of a single subgiant due to magnetic fields that lower convective efficiency and produce large star spots, akin to the effects in young stars noted above, concluding that all of these pathways are possible to some degree. 

Progress on understanding SSGs/RSSs has been hampered by the lack of knowledge in most cases of basic SSG/RSS properties (masses, radii) or even whether they are binaries at all in some cases.
An SSG/RSS comprising an EB would be extremely valuable for helping to resolve these questions by directly linking evolutionary properties to fundamental physical properties. 

In this paper, we report the discovery of \eb\ as an RSS EB, comprising two stars near the base of the red giant branch, and having nearly identical masses, in a 33.9-d orbit. 
The features of this system lend themselves to a detailed, quantitative assessment of physical scenarios that may be representative of the SSG/RSS class. 

In Section~\ref{sec:eb}, we first describe the means by which the \eb\ system was identified and summarize its key observational properties, and Section~\ref{sec:data} presents the data used in our analysis. 
We present a detailed analysis of the spectroscopic, radial-velocity, light-curve, and photometric data in Section~\ref{sec:analysis}, with which we report the fundamental physical properties of the system in Section~\ref{sec:results}. 
We discuss the implications of the observed properties of the \eb\ system for models of SSG/RSS evolutionary pathways in Section~\ref{sec:discussion}, and we conclude with a summary of our findings in Section~\ref{sec:summary}.

\section{The 2M0056--08 System: Identification and Summary of Basic Properties}\label{sec:eb}

\begin{figure}[!t]
    \centering
\includegraphics[width=\linewidth,trim=2 0 7 6,clip]{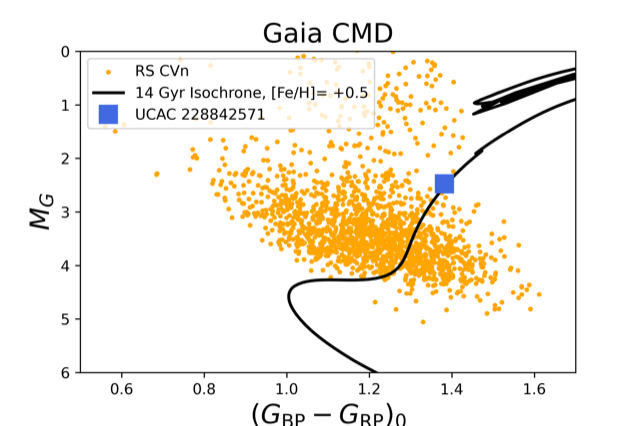}
    \caption{Gaia color-magnitude diagram of RS CVn systems identified by \citet{Leiner:2022}. The \eb\ system is represented by the large symbol; its position can only be accommodated by a normal subgiant if simultaneously extremely old and metal-rich, as indicated by the model isochrone (black curve).}
    \label{fig:cmd}
\end{figure}

\eb\ was identified by \cite{Leiner:2022} in their search for candidate SSGs in the field. 
Its position in the Gaia color-magnitude diagram can only be accommodated by a normal giant star if its age is $\gtrsim$14~Gyr and its metallicity $\gtrsim$0.5 (see Figure~\ref{fig:cmd}). It is also a known X-ray source, which would be quite unusual for a normal giant or subgiant.

The system has a precise parallax in Gaia DR3, placing it at a distance of 677$\pm$9~pc, after application of the small parallax offset reported by \citet{Lindegren:2021} and the small uncertainty inflation factor reported by \citet{ElBadry:2021}. 

\eb\ has a number of other identifiers in the available catalogs, including 
UCAC2~28842571,
SDSS~J005631.83$-$080905.5,
TIC~408036613, and others. In this paper we adopt the shorthand \eb, based on the identifier 
2MASS~J00563182$-$0809055.

\section{Data}\label{sec:data}

\subsection{TESS Light Curve Analysis}\label{subsec:LC}
\begin{figure*}[!ht]
    \centering
    \includegraphics[width=0.65\linewidth,trim=8 0 8 0,clip]{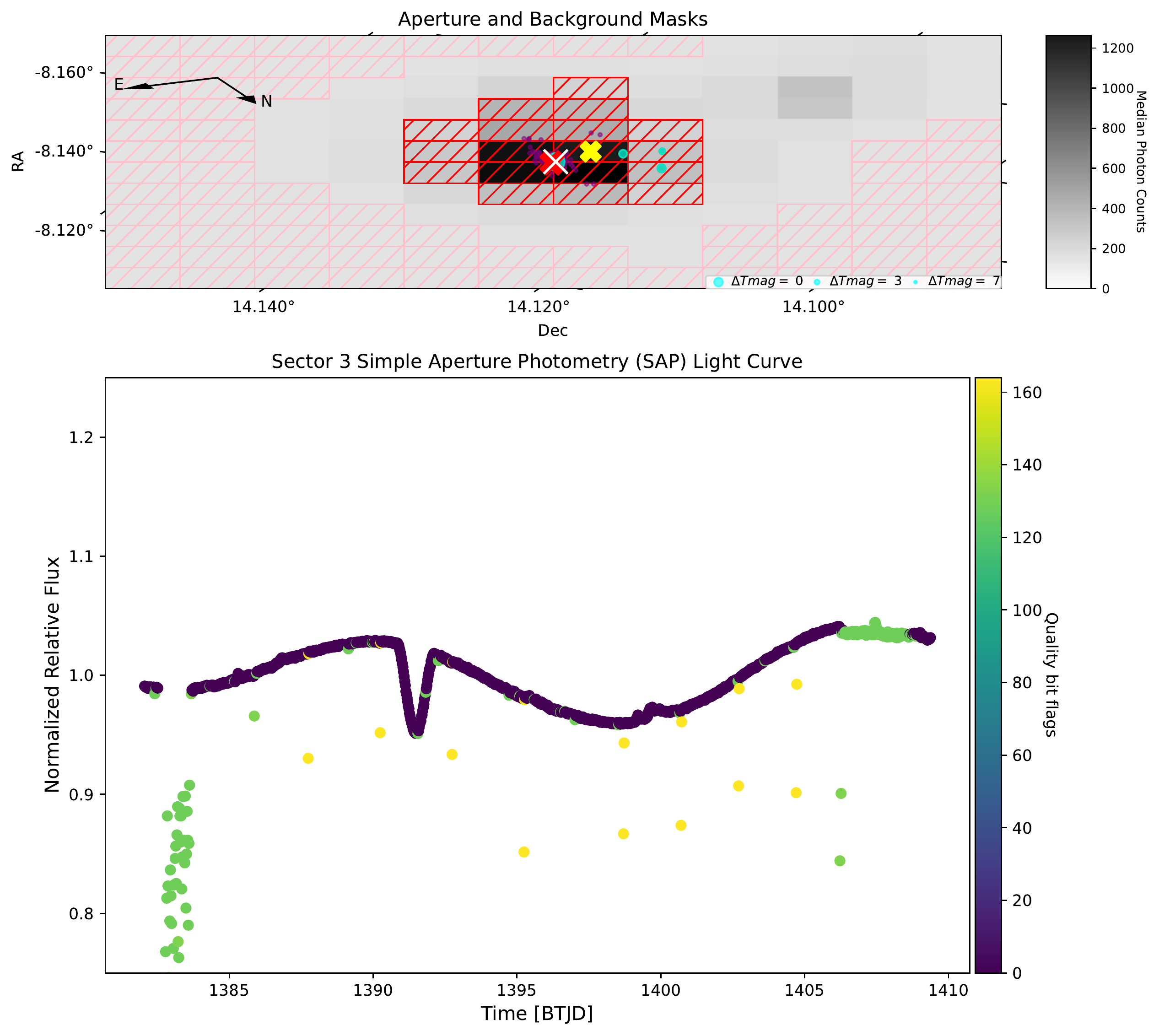}
    \includegraphics[width=0.65\linewidth,trim=8 0 8 0,clip]{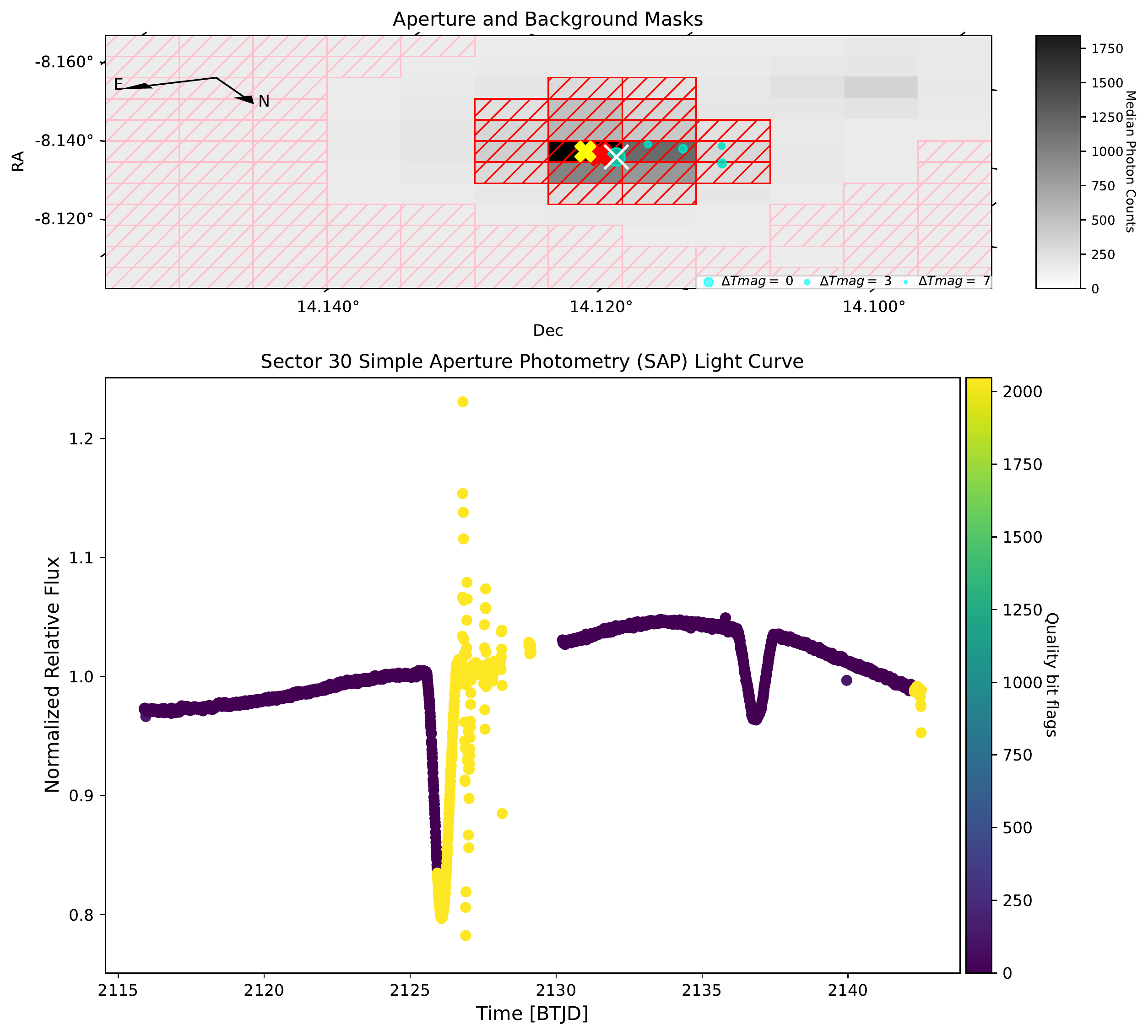}
    \caption{TESS light curve extraction pixel masks and flagged light curve data for Sector~3 (top) and Sector~30 (bottom). The pink and red dashed boxes in the upper panels represent the pixel masks for the sky background and the aperture mask used for photometric extraction with the NEMESIS pipeline. The red and yellow Xs represent the location of the target star and the photocenter of the aperture mask. The cyan colored points are other stellar sources within 42\arcsec~of~\eb, sized by the difference in magnitude from \eb. The lower panels display the light curves with Barycentric TESS Julian Day (BTJD) timestamps in noramlized relative flux units. The photometric measurements are colored by the TESS quality flag bit values for each timestamp. In Sector 30, about half of the deeper eclipse have quality flag bit values of 2048.
    } 
    \label{fig:lc}
\end{figure*}

\eb~(TIC 408036613) was observed by TESS during Sector 3 of cycle 1 (2018-Sep-20 to 2018-Oct-18) in camera 1 and again during Sector 30 of cycle 3 (2020-Sep-22 to 2020-Oct-21) in camera 1. The Full Frame Images (FFI) from Sectors 3 and 30 were downloaded from the Barbara A. Mikulski Archive for Space Telescopes (MAST).
The 30-minute cadence and 10-minute cadence FFI data were then processed using the NEMESIS pipeline \citep{Feliz2021} to produce Simple Aperture Photometry (SAP) light curves (Figure~\ref{fig:lc}).

To correct for flux contamination of other stellar sources within 42\arcsec~of~\eb, we queried the TESS Input Catalog \citep[TIC;][]{StassunTIC:2019} and calculated a flux contamination ratio of $\sim$1.36\% for both Sectors 3 and 30. This value was subtracted from the SAP light curves and then they were normalized.
We detrended the light curves manually, by masking out the eclipses and fitting a cubic spline function to the out-of-eclipse portions in order to remove the variability and then manually removing outliers.

\eb\ has also been observed since 2012 by the All-Sky Automated Survey for
Supernovae project \citep[ASAS-SN;][]{Shappee:2014, Kochanek:2017},
first in the Johnson $V$ band and later in Sloan $g$.\footnote{Data available at \url{https://asas-sn.osu.edu/}}
These data are less precise than the TESS observations, but cover a much
longer interval and capture many more primary and secondary eclipses,
although with sparser coverage. {We have chosen to incorporate
the ASAS-SN light curves into our analysis for their value for 
improving the orbital period. We detrended them in the same
way as the TESS data, and removed several outliers. We retained
1224 measurements in $V$, and 2790 in $g$.}

\subsection{Spectroscopy}
\label{subsec:rv}

Time-series spectroscopy of \eb\ was obtained using the Coud{\'e} spectrograph on the McDonald Observatory Harlan J.\ Smith 2.7-m telescope. The Robert G.\ Tull Coud{\'e} is a cross-dispersed echelle covering 3400–-10000~\AA\ with a resolving power of $R\sim60,000$ using the 1$\farcs$2 slit \citep{Tulletal1995}. Fifteen epochs were obtained between 2020 August and 2021 October. Individual exposure times were 1500~s. 
Wavelength calibration made use of periodic ThAr comparison lamp spectra taken periodically throughout each night. 

Followup spectroscopy was also obtained on the IGRINS spectrograph on Gemini South. IGRINS is an immersion grating echelle spectrograph operating in the $H$ and $K$ bands (1.45--2.5~$\mu$m) at a spectral resolving power of $R\sim45,000$ \citep{Maceetal2016,Maceetal2018}. Nine observations were made between 2021 August and 2022 January. IGRINS spectra are reduced with a tailored pipeline \citep{leeetal2017} performing telluric correction with an A0 spectroscopic standard and wavelength calibration from OH emission and telluric absorption lines. 


\begin{figure}[!ht]
    \centering
    \includegraphics[width=\linewidth,trim=0 0 0 30,clip]{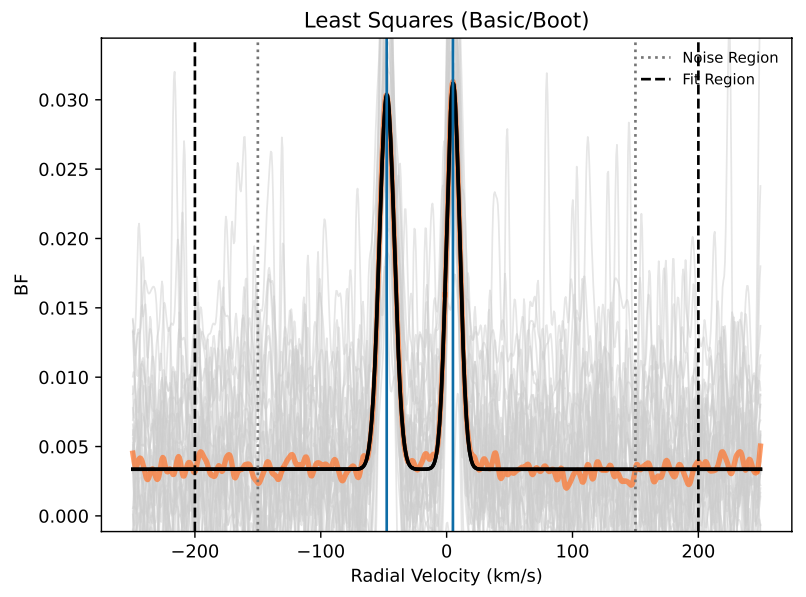}
    \caption{Broadening function of a representative spectroscopic observation of \eb, showing the two eclipsing components to be of comparable brightness in visible light. Note also the lack of any clear signature of a tertiary companion; its contribution is constrained to be less than a few percent of the total system light in the visible.}
    \label{fig:lsd}
\end{figure}

\subsubsection{Radial Velocities}
\label{sec:rvs}

We derive radial velocities from the Coud{\'e} and IGRINS spectra by computing spectral-line broadening functions \citep[BFs;][]{Rucinski1992,Tofflemire:2019}. The BF results from the linear inversion of an observed spectrum with a narrow-lined template, and represents a reconstruction of the absorption-line profile in an observed spectrum as a function of radial velocity (RV). If the observed spectrum contains light from two stars, as it does in \eb, two profiles emerge in the BF, each at their individual RVs. Figure~\ref{fig:lsd} presents an example BF from a Coud{\'e} spectrum.

\begin{figure}[!t]
    \centering
    \includegraphics[width=\linewidth,trim=0 0 70 235,clip]{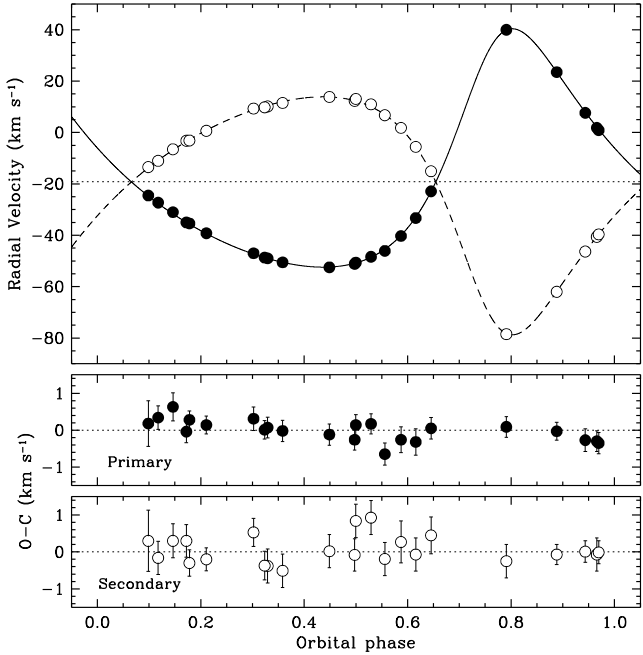}
    \caption{Radial-velocity measurements for \eb\ together with our binary
    model described below. Residuals are shown at the bottom.
    {The errorbars correspond to the final uncertainties,
    after folding in the jitter values reported in Table~\ref{tab:mcmc},
    added in quadrature.}
}
    \label{fig:orbit}
\end{figure}

We use a synthetic model spectrum from the \citet{husser2013} PHOENIX suite as our narrow-lined template with a \teff\ of {4100~K} and $\log g$ of {3.5}. BFs are computed for individual spectral orders that are devoid of heavy telluric contamination and then combined, weighted by their baseline signal-to-noise. This high S/N BF is then fit with a rotational-broadened line profile \citep{Gray2008} for each component to measure their RVs. Uncertainty on the RVs are computed with a bootstrap approach, where $10^5$ combined BFs are constructed and fit, sampling the contributing orders randomly with replacement. The standard deviation of the resultant RV distribution corresponds to our RV error. Our RV measurements are reported in Table~\ref{tab:rvs}, and shown graphically in Figure~\ref{fig:orbit} together with our best-fit spectroscopic orbit model described later.

For the Coud{\'e} spectra, 34 orders were used spanning $\sim$4800--8850~\AA. Many of our Coud{\'e} epochs consisted of three individual spectra taken back to back. In these cases, we analyze each spectrum individually and report the error-weighted mean and uncertainty. 

For the IGRINS spectra, we only analyze $H$-band orders due to an RV offset between the $H$ and $K$ bands, and because the $H$ band has a higher spectral information content. Fifteen orders are used across the $H$ band spanning $\sim$15200--17300~\AA.

\subsubsection{Projected Rotation Velocity}\label{subsec:vsini}

As a representation of the average absorption line profile, BFs are also amenable to measurements of the projected stellar rotation velocity (\vsini). In addition to rotational broadening, the lines are also broadened by macroturbulence and the instrument's spectral resolution. (Microturbulent broadening is included in the synthetic template.) The BFs are not sufficiently broad to simultaneously fit the \vsini\ and macroturbulent velocity (\vmac), so we adopt the temperature dependent empirical relation from \citet{ValentiFischer2005} with the systematic offset applied from \citet{Doyleetal2014}. For the measured \teff\ values we compute \vmac\ values of 1.9 and 2.1 \kms\ for the primary and secondary, respectively. Setting these values, we fit rotationally-broadened profiles to the Coud{\'e} spectra where the BFs do not overlap. Errors are assessed as described in Section~\ref{sec:rvs}. Subtracting the instrumental broadening in quadrature and taking the error-weighted mean and standard deviation we measure \vsini\ values of ${7.6}\pm0.2$~\kms\ and ${6.5}\pm{0.4}$~\kms\ for the primary and secondary, respectively. 


\begin{deluxetable}{lccl}
\tablecaption{RV Measurements for \eb \label{tab:rvs}}
\tablehead{
\colhead{BJD} & 
\colhead{$v_1$} & 
\colhead{$v_2$} &
\colhead{Inst.}
\\
\colhead{(2,400,000+)} & 
\colhead{(\kms)} & 
\colhead{(\kms)} &
\colhead{}
}
\startdata
    59067.9228  &  \phn\phs$1.80 \pm 0.08$   &  $-40.62 \pm 0.08$        &  Coud\'e \\
    59095.8756  &  \phs$39.94 \pm 0.04$      &  $-78.48 \pm 0.03$        &  Coud\'e \\
    59108.7912  &  $-35.02 \pm 0.13$         &  \phn$-3.23 \pm 0.01$     &  Coud\'e \\
    59119.8170  &  $-51.15 \pm 0.08$         &  \phs$12.28 \pm 0.01$     &  Coud\'e \\
    59121.7938  &  $-46.09 \pm 0.13$         &  \phn\phs$6.72 \pm 0.04$  &  Coud\'e \\
    59122.8517  &  $-40.30 \pm 0.22$         &  \phn\phs$1.79 \pm 0.36$  &  Coud\'e \\
    59123.8195  &  $-33.29 \pm 0.24$         &  \phn$-5.60 \pm 0.07$     &  Coud\'e \\
    59124.8223  &  $-22.91 \pm 0.10$         &  $-15.08 \pm 0.24$        &  Coud\'e \\
    59174.7040  &  $-27.31 \pm 0.17$         &  $-11.01 \pm 0.08$        &  Coud\'e \\
    59175.6711  &  $-31.03 \pm 0.27$         &  \phn$-6.54 \pm 0.11$     &  Coud\'e \\
    59187.6528  &  $-50.63 \pm 0.05$         &  \phs$13.07 \pm 0.09$     &  Coud\'e \\
    59188.6453  &  $-48.39 \pm 0.07$         &  \phs$10.96 \pm 0.11$     &  Coud\'e \\
    59439.8420  &  \phn\phs$7.70 \pm 0.27$   &  $-46.25 \pm 0.25$        &  IGRINS \\
    59440.7184  &  \phn\phs$0.92 \pm 0.25$   &  $-39.59 \pm 0.28$        &  IGRINS \\
    59452.9127  &  $-48.98 \pm 0.07$         &  \phs$10.13 \pm 0.10$     &  Coud\'e \\
    59453.8904  &  $-50.55 \pm 0.09$         &  \phs$11.48 \pm 0.09$     &  Coud\'e \\
    59477.5961\tablenotemark{a} &  $-16.29 \pm 0.63$         &  $-22.06 \pm 0.42$        &  IGRINS \\
    59481.6669  &  $-35.35 \pm 0.19$         &  \phn$-3.05 \pm 0.33$     &  IGRINS \\
    59485.8678  &  $-46.94 \pm 0.29$         &  \phs\phn$9.38 \pm 0.35$  &  IGRINS \\
    59486.5909  &  $-48.62 \pm 0.20$         &  \phs\phn$9.86 \pm 0.36$  &  IGRINS \\
    59490.8374  &  $-52.48 \pm 0.10$         &  \phs$13.84 \pm 0.06$     &  Coud\'e \\
    59505.7283  &  \phs$23.54 \pm 0.19$      &  $-61.91 \pm 0.23$        &  IGRINS \\
    59516.6548  &  $-39.15 \pm 0.19$         &  \phn\phs$0.71 \pm 0.28$  &  IGRINS \\
    59580.6096  &  $-24.47 \pm 0.60$         &  $-13.40 \pm 0.82$        &  IGRINS \\
\enddata
\tablenotetext{a}{Observation excluded from the analysis in Section~\ref{subsec:lcfit} for showing large unexplained residuals.}
\end{deluxetable}

\subsubsection{Spectroscopic Flux Ratio}
\label{sec:fluxratios}

In addition to measuring the overall velocity of each star across the entire spectrum, we also analyze each order for all of the epochs independently in order to examine the flux contribution of each star as a function of wavelength. For this purpose, we compute the cross-correlation function (CCF) by cross-correlating the spectrum in a given order against a high-resolution template with \teff\ of {4100 K} and $\log g$ of {3.5} from PHOENIX \citep{husser2013}, using PyXCSAO \citep{rvsao,pyxcsao}, which enables filtering out the low spatial Fourier frequencies. Afterwards, the resulting CCF was fit with two Gaussian functions.

Not all of the orders have a sufficient number of spectral lines and/or sufficiently high signal-to-noise to achieve meaningful cross-correlation. We select only those orders where the velocity of each component measured in a given order was consistent with the velocity measured across the entire spectrum to within 5 \kms, the full width at half maximum of both of the fitted Gaussians was between 6 and 16 \kms, and the amplitude of each component is $>$ 0. Of the total 2080 orders across all of the epochs, only 440 satisfied these requirements.

We then integrate under the Gaussian-fitted CCF profile for each star, and compare the ratio of their respective areas. The resulting value is representative of the flux ratio in the system,
provided the temperature difference between the components is sufficiently small (see below). 
To determine the mean flux ratio in each of the bandpasses relevant for our analysis, we {first} averaged all of the individual measurements from spectral orders that fall within the wavelength range of each filter, {epoch by epoch, and then took the mean over all epochs.} The resulting flux ratios $\ell_2/\ell_1$ for the $g$, $V$, TESS, and $H$ bandpasses are, respectively, ${0.851 \pm 0.024}$, ${0.875 \pm 0.026}$, ${0.848 \pm 0.019}$, and ${0.696 \pm 0.041}$. The uncertainties are formal only, and correspond to the error of the mean. They do not include possible systematic errors that are difficult to quantify, such as variability due to magnetic activity (spots).

Another bias may be caused by the fact that the stars do not have the same temperature, and therefore differ in their intrinsic line strengths. {For this reason, the values reported above should be considered strictly as line flux ratios, rather than continuum flux ratios. To explore the difference, we carried out numerical simulations to recreate our measuring process. We generated synthetic double-lined spectra by combining templates appropriate for the two stars with temperatures of 4100 and 4300~K for the primary and secondary, and with a range of Doppler shifts reflecting the true velocity differences measured in the real spectra. The templates were added together in a proportion near the average of the flux ratio estimates given above ($\sim$0.85). We then cross-correlated these synthetic double-lined spectra against the template for one of the stars, and measured the ratio of the areas of the CCF peaks in the same way as done with the observed spectra, order by order and epoch by epoch. Next, we averaged the results for the orders within each bandpass at each epoch, and finally we calculated the mean over all epochs. 
We find these synthetic line flux ratios to be about 2\% smaller than the input (continuum) flux ratio of 0.85 that we used to generate the artificial double-lined spectra. This indicates that our measured flux ratios from the real spectra are likely to also be lower than the true continuum ratio. Dividing the measured line flux ratios from our simulations by the input continuum ratio gives $0.981 \pm 0.005$ ($g$), $0.978 \pm 0.003$ ($V$), $0.983 \pm 0.006$ (TESS), and $0.982 \pm 0.011$ ($H$). These factors may be used to convert the measured line flux ratios in the previous paragraph to continuum flux ratios. In this way we obtain final values for $\ell_2/\ell_1$ of $0.867 \pm 0.025$ ($g$), $0.895 \pm 0.027$ ($V$), $0.863 \pm 0.020$ (TESS), and $0.709 \pm 0.043$ ($H$). As before, these are formal uncertainties that do not account for the impact of stellar activity.}

%

Although there is considerable scatter in the individual measurements {from the real spectra} across all epochs, there {is a clear trend} in the flux ratio as a function of $\lambda$: from $\sim$0.9 at 4000\AA\ in the Coud{\'e} data to $\sim$0.7 at 1.6~$\mu$m in IGRINS. Such a trend would be consistent with a system in which the cooler star has the larger radius.
{We note also that the} Coud{\'e} data from late 2020 (10 epochs in total) deviate somewhat from this trend, favoring a smaller flux ratio of $\sim$0.65 at visible wavelengths. 
{The} time when these data were obtained corresponds to {a flux minimum within} the long-term photometric variability {cycle} for the system (see Section~\ref{sec:continuum}), whereas {most of} the other Coud{\'e} epochs as well as all of the IGRINS observations were conducted across a more representative range of the long-term photometric variation. {As the TESS observations, which have limited time coverage, were all gathered at times when the brightness was more or less average, we chose not to exclude any of the Coud{\'e} spectra for computing the mean flux ratio in the TESS band.}




\begin{figure}[!t]
    \centering
    \includegraphics[width=\linewidth]{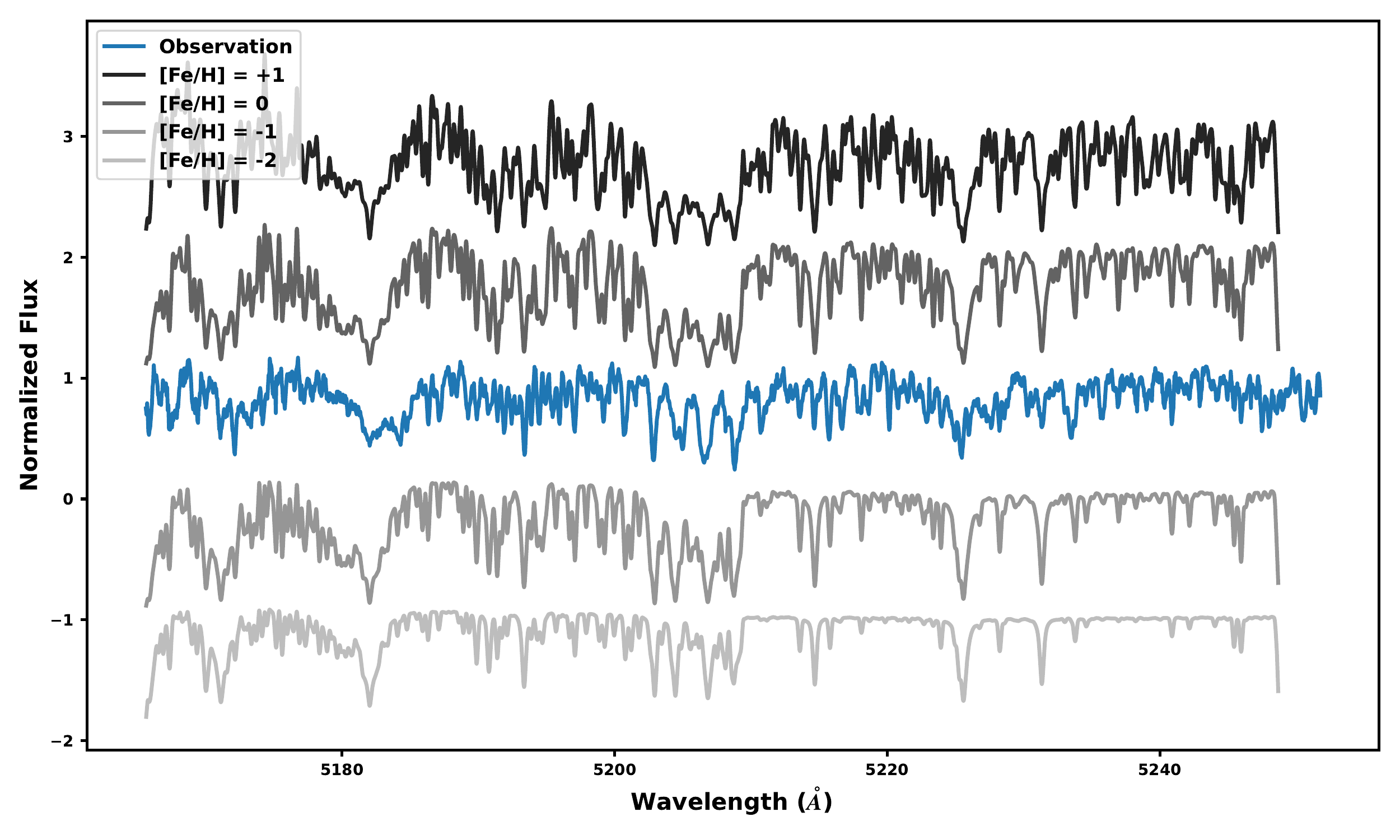}
    \caption{Comparison of the observed Coud\'e spectrum of \eb\ (blue) with synthetic binary spectra at a range of metallicities.}
    \label{fig:metal}
\end{figure}

\subsubsection{Metallicity}\label{subsec:metallicity}

To estimate the metallicity of the \eb\ system, we used the highest signal-to-noise ratio Coud\'e spectrum from an epoch when the eclipsing stars were well separated in radial velocity, and compared it to synthetic spectra created from two PHOENIX models corresponding to the same \teff, $\log g$, and radial-velocity separation, as shown in Figure~\ref{fig:metal}. From this comparison, we conclude that the system has a metallicity [M/H] between 0 and $-1$. Hence we adopt an estimate of $-0.5 \pm 0.5$.

\subsubsection{Chromospheric and Coronal Activity}\label{subsec:activity}

We observe clear emission from both stars in the chromospheric activity-sensitive lines of Ca H \& K (not shown here).
Chromospheric activity is rare among evolved stars because stars generally slow their spins considerably as their radii increase on the subgiant and red giant branches. However, a small minority of red giants are known to rotate rapidly and consequently exhibit chromospheric activity. \citet{Dixon:2020} used a large sample of rapidly rotating red giants observed by SDSS/APOGEE to develop an empirical rotation-activity relation for evolved stars, akin to the long-established rotation-activity relationship for cool stars on the main sequence. Given the \vsini\ that we measure for \eb, the \citet{Dixon:2020} relations predict a UV excess in the {\it GALEX\/} NUV band of $\approx$1.1~mag, which largely explains the NUV excess clearly seen in the SED of \eb\ (Figure~\ref{fig:sed}). 


\eb\ is also an X-ray source in the ROSAT catalog, with a total flux of $1.18 \pm 0.42 \times 10^{-12}$ erg~s$^{-1}$~cm$^{-2}$ and a hardness ratio of 0.95, which we represent in Figure~\ref{fig:sed} as a hard and soft component (nearly all of the flux is in the hard component due to the extremely high hardness ratio). The slope of the X-ray emission is consistent with a hot corona having a temperature of $\sim$5 million K, represented as a simple blackbody in Figure~\ref{fig:sed} by the magenta curve through the X-ray fluxes. (The blackbody has been extincted by the same $A_V$ as inferred from the overall SED.) 

\citet{Richey-Yowell:2019} used observations of active cool stars to develop empirical relationships between chromospheric emission in the NUV and FUV versus coronal emission in X-rays, finding that the X-ray flux is generally comparable in energy to the NUV excess, as observed in Figure~\ref{fig:sed}. Those same relations predict that the chromospheric emission produces an FUV excess that is $\lesssim$10\% of the NUV excess. We correct the NUV and FUV fluxes accordingly for the chromospheric contribution in Figure~\ref{fig:sed}; the FUV correction is negligible, leaving a significant FUV excess. We return to a discussion of the possible origins of the large FUV excess in Section~\ref{subsec:sed2}.

\section{Analysis}\label{sec:analysis}

\subsection{Initial Constraints on Stellar Properties}\label{subsec:sed1}

\begin{figure*}[!t]
    \centering
    \includegraphics[width=0.98\linewidth,trim=95 70 50 50,clip]{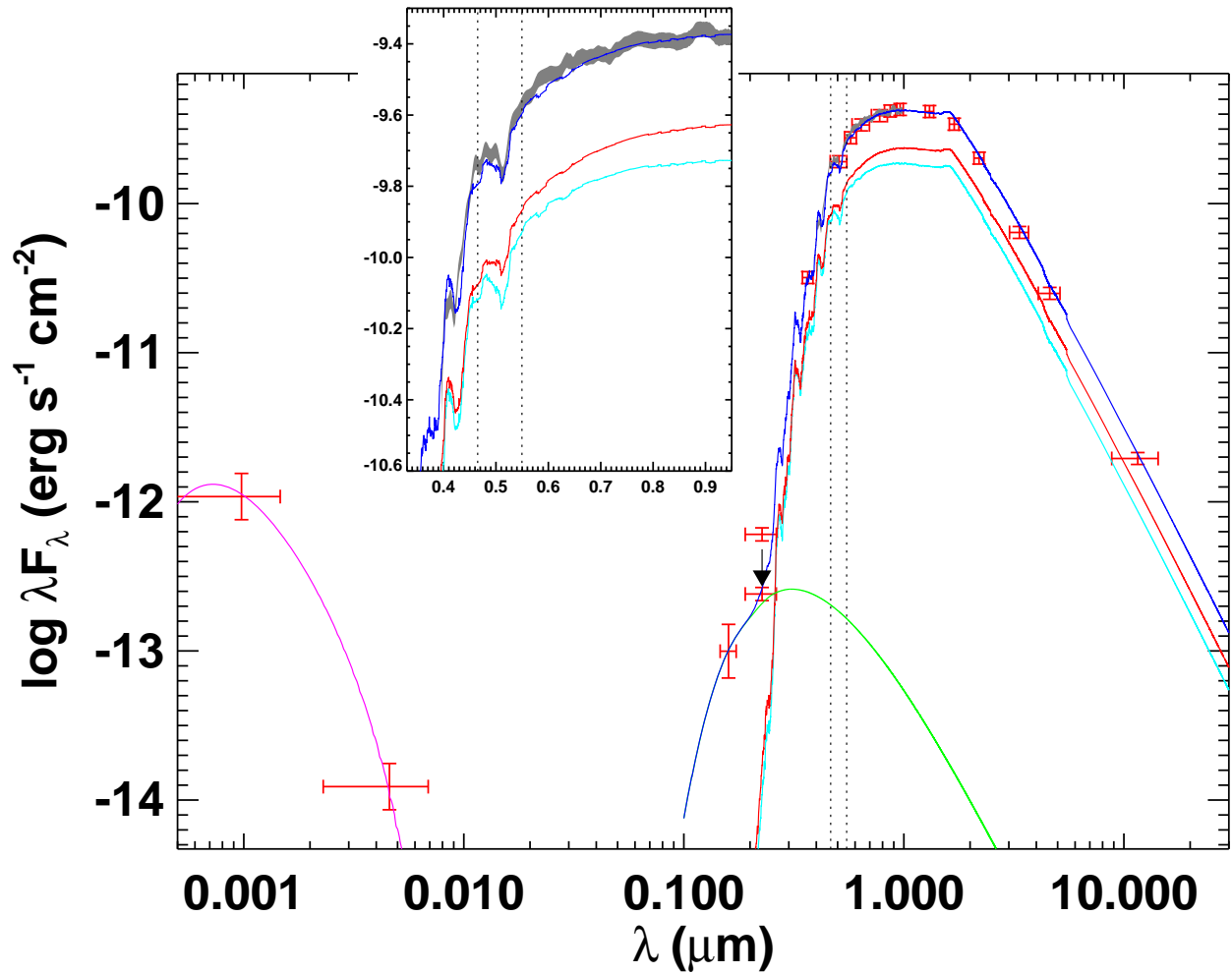}
    \caption{Spectral energy distribution of \eb. Red symbols represent the observed broadband photometry from ROSAT, GALEX, SDSS, Pan-STARRS, 2MASS, and WISE; horizontal bars represent the effective bandpass widths. Red and cyan curves represent PHOENIX stellar atmosphere models representing the cooler and hotter EB component stars, respectively, and the green curve represents a hot blackbody ($T = 11\,000$~K) with $R = 0.03$~\rsun, chosen to match the (activity-corrected; downward arrow) GALEX fluxes; see the text. The dark blue curve represents the sum of all three components, providing an excellent fit to the broadband photometry as well as to the Gaia spectrum (gray swath, inset), which also constrains the metallicity via the absorption feature in the $g$ band (vertical dotted lines). The magenta curve represents a blackbody with best-fit $T = 5\times 10^6$~K to the X-ray fluxes, consistent with coronal emission from the active eclipsing stars; see the text.}
    \label{fig:sed}
\end{figure*}

In order to obtain an initial estimate of the components' effective temperatures (\teff), we performed a fit to the combined-light, broadband SED of the \eb\ system (Figure~\ref{fig:sed}), including broadband photometry spanning the wavelength range 0.15--22~$\mu$m from GALEX, SDSS, Pan-STARRS, {\it Gaia\/} DR3, 2MASS, and WISE \citep[see][for details of the SED fitting methodology specifically in the context of EBs]{StassunTorres:2016,Miller:2020}. 

We started by treating the SED as a single star with $T_{\rm eff} \approx 4250$~K based on the value reported in the TIC and adopting the maximum line-of-sight extinction of $A_V = 0.22$ from the Galactic dust maps of \citet{Schlegel:1998} given the system's distance and Galactic latitude. 
The fact that the SED can be reasonably approximately by a single-star fit (not including the clear UV excess in the GALEX data; see Section~\ref{subsec:sed2}), suggests that the two eclipsing components have \teff\ comparable to one another and close to 4250~K. This is reinforced by the similarity of the peaks in the spectral broadening function (Figure~\ref{fig:lsd}) and of the primary and secondary eclipses in the light curve data, though that they are not identical clearly indicates that the two components must have slightly different \teff. 
We perform a complete, multi-component fit to the SED in Section~\ref{subsec:sed2}, but utilize the above \teff\ estimate in our light-curve analysis, as we now discuss.

\subsection{Spectroscopic and Light Curve Analysis}\label{subsec:lcfit}

Modeling of the TESS 30\,m and 2\,m cadence photometry was performed
simultaneously with the ASAS-SN $V$-band and $g$-band photometry and
with the radial velocities of the primary and secondary from the Coud\'e
and IGRINS spectrographs. The analysis was carried out within a Markov
chain Monte Carlo (MCMC) framework using the {\tt eb} code of
\cite{Irwin:2011}, which is based on the Nelson-Davis-Etzel binary
model \citep{Etzel:1981, Popper:1981} that is well suited to wide
systems with nearly spherical components such as those of \eb.

The standard light curve parameters we solved for are the orbital
period ($P$), time of primary eclipse ($T_0$), the central
surface brightness ratio between the secondary and primary in each
bandpass ($J \equiv J_2/J_1$, $J_{\rm TESS}$, $J_V$, $J_g$), the sum
of the relative radii normalized by the semimajor axis ($r_1+r_2$),
the radius ratio ($k \equiv r_2/r_1$), the cosine of the orbital
inclination angle ($\cos i$), and the eccentricity ($e$) and argument
of periastron for the primary ($\omega_1$) expressed in terms of the
variables $\sqrt{e} \cos\omega_1$ and $\sqrt{e}\sin\omega_1$. As a
precaution we allowed for the presence of extra flux in the
photometric aperture by solving for a third light parameter in each
bandpass: $\ell_3$(TESS), $\ell_3$($V$), and $\ell_3$($g$). Third light is defined here such
that $\ell_1 + \ell_2 + \ell_3 = 1$, where $\ell_1$ and $\ell_2$ for
this normalization are taken to be the light at first quadrature.

The spectroscopic parameters in our analysis were the velocity
semiamplitudes of the primary and secondary ($K_1$, $K_2$) and the
center-of-mass velocity of the system ($\gamma$). To allow for the
possibility that the velocity zero-points of the Coud\'e and IGRINS
spectrographs may be different, we solved for a separate $\gamma$
value for each instrument ($\gamma_{\rm Coud\acute{e}}$, $\gamma_{\rm
  IGRINS}$).

{The detrending and normalization of the photometry artificially
remove variability out of eclipse. While effects due to gravity darkening and reflection may still be present during the eclipses, both are expected to be very small for this system by virtue of the long orbital period and the related fact that the stars are nearly spherical (see below). Furthermore, any real photometric variations from those causes are likely to be overwhelmed by distortions in the light curve due to spots. For these reasons, we have chosen here to
use} the option in {\tt eb} that suppresses those effects in computing
the binary model. {Additionally, as the flattened out-of-eclipse portions of the light curve contribute no information, we restricted our analysis of the TESS data to regions within 0.1 in phase from the center of each eclipse, corresponding to approximately 2.5 times the width of the eclipses.} For the TESS photometry in sector~3 (full-frame
images) we oversampled the model light curve at each iteration of our
solution, and then integrated over the 30\,m duration of each cadence
prior to the comparison with the observations
\citep[see][]{Gilliland:2010, Kipping:2010}.

Limb darkening was treated slightly differently for TESS and ASAS-SN.
For TESS we adopted a linear limb-darkening law, and allowed the
coefficient to vary independently for each component ($u_1$, $u_2$).
Experiments with a quadratic law brought no improvement, and made
convergence more difficult given the already large number of free
parameters. The sparser coverage and poorer precision of the ASAS-SN
photometry did not allow us to solve for the linear limb-darkening
coefficients, so those quantities were set to their theoretically
expected values according to the tabulation of \cite{Claret:2011},
based on the stellar properties. In the $V$ band the coefficients
adopted are {0.822 and 0.812} for the primary and secondary, and for the Sloan $g$ band they are {0.903 and 0.889}.

Weights for the various data sets were based on the observational
uncertainties, modified by including adjustment parameters that we
solved for simultaneously and self-consistently with the other
variables in our MCMC analysis, following \cite{Gregory:2005}. For the
photometry we included multiplicative scale factors for the internal
errors, as such errors can sometimes be under- or overestimated. We did
this separately for each ASAS-SN bandpass and separately also for the
TESS photometry from sectors~3 and 30 (four scale factors). For the
radial velocities, we solved for ``jitter'' terms added quadratically
to the internal errors, independently for each star and each
instrument (four jitter values).

The MCMC analysis was carried out using the {\tt
  emcee\/}\footnote{\url{https://github.com/dfm/emcee}} code of
\cite{Foreman-Mackey:2013}. The total number of adjustable parameters
is 27.  We used 100 walkers with 30000 links each, after discarding
the burn-in. We adopted uniform or log-uniform priors over suitable
ranges for all variables, as listed below in
Table~\ref{tab:mcmc}. Convergence was verified by visual
examination of the chains, and by the requirement of a Gelman-Rubin
statistic of 1.05 or smaller for each parameter \citep{Gelman:1992}.

As is commonly found in eclipsing systems with similar components and
partial eclipses, the radius ratio in our initial solutions was found
to be poorly constrained by the data, partly because of the poor
coverage of the egress of the deeper eclipse. In such cases it is
helpful to impose an external constraint on the flux ratio, because
the flux ratio depends very strongly on the radius ratio:
$\ell_2/\ell_1 \propto k^2$. Our spectroscopic observations described
in Section~\ref{subsec:rv} provide the necessary constraints,
which we applied as Gaussian priors in our solution, in each of the three bandpasses (TESS, $V$, $g$).

\section{Results}\label{sec:results}

\setlength{\tabcolsep}{5pt}
\begin{deluxetable}{lcc}
\tablewidth{1.0\columnwidth}
\tablecaption{Joint Photometric-Spectroscopic Orbital Solution for \eb \label{tab:mcmc}}
\tablehead{ \colhead{~~~~~~~Parameter~~~~~~~} & \colhead{Value} & \colhead{Prior} }
\startdata
 $P$     (days)               &  $33.879495^{+0.000051}_{-0.000049}$   & [30, 35] \\ [1ex]
 $T_0$ (BJD$-$2,400,000)      &  $59136.8475^{+0.0012}_{-0.0012}$   & [59130, 59140] \\ [1ex]
 $\gamma_{\rm Coud\acute{e}}$ (\kms)  &  $-19.240^{+0.064}_{-0.062}$         & [$-$30, 0]  \\ [1ex]
 $\gamma_{\rm IGRINS}$ (\kms)         &  $-19.184^{+0.077}_{-0.078}$         & [$-$30, 0]  \\ [1ex]
 $K_1$ (\kms)                 &  $46.370^{+0.101}_{-0.095}$            & [30, 60]   \\ [1ex]
 $K_2$ (\kms)                 &  $46.45^{+0.14}_{-0.14}$               & [30, 60]   \\ [1ex]
 $\sqrt{e} \cos\omega_1$      &  $+0.47854^{+0.00069}_{-0.00069}$      & [$-$1, 1] \\ [1ex]
 $\sqrt{e} \sin\omega_1$      &  $-0.3510^{+0.0020}_{-0.0020}$         & [$-$1, 1] \\ [1ex]
 $r_1+r_2$                    &  $0.13772^{+0.00091}_{-0.00092}$       & [0.01, 0.50] \\ [1ex]
 $k \equiv r_2/r_1$           &  $0.844^{+0.011}_{-0.013}$          & [0.5, 2.0] \\ [1ex]
 $\cos i$                     &  $0.0843^{+0.0012}_{-0.0013}$          & [0, 1] \\ [1ex]
 $J_{\rm TESS}$               &  $1.013^{+0.020}_{-0.020}$             & [0.3, 2.5] \\ [1ex]
 $J_{\rm V}$                  &  $1.287^{+0.049}_{-0.043}$             & [0.3, 2.5] \\ [1ex]
 $J_{\rm g}$                  &  $1.227^{+0.040}_{-0.035}$             & [0.3, 2.5] \\ [1ex]
 $\ell_3$, TESS               &  $0.172^{+0.017}_{-0.015}$             & [0.0, 0.5] \\ [1ex]
 $\ell_3$, $V$                &  $0.039^{+0.029}_{-0.023}$             & [0.0, 0.5] \\ [1ex]
 $\ell_3$, $g$                &  $0.022^{+0.025}_{-0.016}$             & [0.0, 0.5] \\ [1ex]
 $u_1$, TESS                  &  $0.830^{+0.046}_{-0.044}$             & [0.0, 1.0] \\ [1ex]
 $u_2$, TESS                  &  $0.450^{+0.082}_{-0.086}$             & [0.0, 1.0] \\ [1ex]
 $\sigma_1$, Coud\'e (\kms)   &  $0.234^{+0.069}_{-0.051}$             & [0, 3] \\ [1ex]
 $\sigma_2$, Coud\'e (\kms)   &  $0.413^{+0.103}_{-0.075}$             & [0, 3] \\ [1ex]
 $\sigma_1$, IGRINS (\kms)    &  $0.103^{+0.119}_{-0.072}$             & [0, 3] \\ [1ex]
 $\sigma_2$, IGRINS (\kms)    &  $0.125^{+0.153}_{-0.089}$             & [0, 3] \\ [1ex]
 $f_{\rm 30\,m}$              &  $6.81^{+0.12}_{-0.12}$             & [$-$5, 5] \\ [1ex]
 $f_{\rm 2\,m}$               &  $0.616^{+0.026}_{-0.024}$             & [$-$5, 5] \\ [1ex]
 $f_{\rm V}$                  &  $1.682^{+0.034}_{-0.034}$             & [$-$5, 5] \\ [1ex]
 $f_{\rm g}$                  &  $1.807^{+0.024}_{-0.024}$             & [$-$5, 5] \\ [1ex]
\enddata
\tablecomments{The values listed correspond to the median of the
  posterior distributions, and the uncertainties represent the 68.3\%
  credible intervals. Priors in square brackets are uniform over the
  specified ranges, except the ones for $f_{\rm 30\,m}$, $f_{\rm 2\,m}$,
  $f_{\rm V}$ and $f_{\rm g}$, which are log-uniform. The limb-darkening coefficients adopted for
  the ASAS-SN $V$ band are {0.822 and 0.812} for the primary and
  secondary. For the ASAS-SN $g$ band they are {0.903 and 0.889},
  respectively.}
\end{deluxetable}
\setlength{\tabcolsep}{6pt}

\subsection{Fundamental Stellar Properties}

The results of our analysis are presented in Table~\ref{tab:mcmc}, and
other properties derived from the adjustable parameters are listed in Table~\ref{tab:derived}.
{As mentioned earlier, the stars in \eb\ are very close to being spherical. Based on our solution, we estimate their oblateness to be {0.00062 and 0.00038}, as defined, e.g., by \cite{Binnendijk:1960}. These values are well below the limit of 0.04 considered to be safe for the Nelson-Davis-Etzel binary model \citep[see][]{Popper:1981}, and support our assumption that gravity darkening effects are negligible for this system.}

\setlength{\tabcolsep}{5pt}
\begin{deluxetable}{lcc}
\tablewidth{1.0\columnwidth}
\tablecaption{Derived Parameters from our Joint Photometric-Spectroscopic Solution for \eb \label{tab:derived}}
\tablehead{ \colhead{~~~~~~~Parameter~~~~~~~} & \colhead{Value} & \colhead{Prior} }
\startdata
 $r_1$                        &  $0.07471^{+0.00039}_{-0.00038}$       & \nodata \\ [1ex]
 $r_2$                        &  $0.06303^{+0.00079}_{-0.00089}$       & \nodata \\ [1ex]
 $i$ (degree)                 &  $85.162^{+0.077}_{-0.067}$            & \nodata \\ [1ex]
 $e$                          &  $0.35224^{+0.00078}_{-0.00077}$       & \nodata \\ [1ex]
 $\omega_1$ (degree)          &  $323.74^{+0.20}_{-0.20}$              & \nodata \\ [1ex]
 $M_1$ ($M_{\sun}$)           &  $1.4202^{+0.0091}_{-0.0089}$          & \nodata \\ [1ex]
 $M_2$ ($M_{\sun}$)           &  $1.4177^{+0.0075}_{-0.0072}$          & \nodata \\ [1ex]
 $q \equiv M_2/M_1$           &  $0.9982^{+0.0037}_{-0.0037}$          & \nodata \\ [1ex]
 $a$ ($R_{\sun}$)             &  $62.38^{+0.12}_{-0.11}$               & \nodata \\ [1ex]
 $R_1$ ($R_{\sun}$)           &  $4.661^{+0.026}_{-0.025}$             & \nodata \\ [1ex]
 $R_2$ ($R_{\sun}$)           &  $3.932^{+0.051}_{-0.057}$              & \nodata \\ [1ex]
 $R_1+R_2$ ($R_{\sun}$)       &  $8.592^{+0.059}_{-0.061}$             & \nodata \\ [1ex]
 $\log g_1$ (cgs)             &  $3.2537^{+0.0047}_{-0.0047}$           & \nodata \\ [1ex]
 $\log g_2$ (cgs)             &  $3.401^{+0.012}_{-0.011}$           & \nodata \\ [1ex]
 $\ell_2/\ell_1$, TESS        &  $0.847^{+0.017}_{-0.017}$             & $G(0.863, 0.020)$ \\ [1ex]
 $\ell_2/\ell_1$, $V$         &  $0.920^{+0.024}_{-0.024}$             & $G(0.895, 0.027)$ \\ [1ex]
 $\ell_2/\ell_1$, $g$         &  $0.879^{+0.021}_{-0.021}$             & $G(0.867, 0.025)$ \\ [1ex]
 $J_{\rm ave}$, TESS          &  $1.190^{+0.033}_{-0.029}$             & \nodata \\ [1ex]
 $J_{\rm ave}$, $V$           &  $1.295^{+0.050}_{-0.043}$             & \nodata \\ [1ex]
 $J_{\rm ave}$, $g$           &  $1.232^{+0.040}_{-0.035}$             & \nodata \\ [1ex]
 $T_{\rm eff,1}$ (K) & $4138 \pm 95$ & \nodata \\ [1ex] 
 $T_{\rm eff,2}$ (K) & $4273 \pm 89$ & \nodata \\ [1ex] 
\enddata
\tablecomments{The values listed correspond to the median of the
  posterior distributions, and the uncertainties represent the 68.3\%
  credible intervals. Priors for the flux ratios are Gaussian, and are indicated as
  $G({\rm mean}, \sigma)$.
  The $J_{\rm ave}$ values are the disk-integrated surface
  brightness ratios, as opposed to the values at the center of the disk presented
  in Table~\ref{tab:mcmc}. The \teff\ and [M/H] values are determined from the SED fitting constrained by the other parameters in this Table; see Section~\ref{subsec:sed2}.}
\end{deluxetable}
\setlength{\tabcolsep}{6pt}

\begin{figure}[!t]
    \centering
    \includegraphics[width=0.95\linewidth,trim=0 0 65 335,clip]{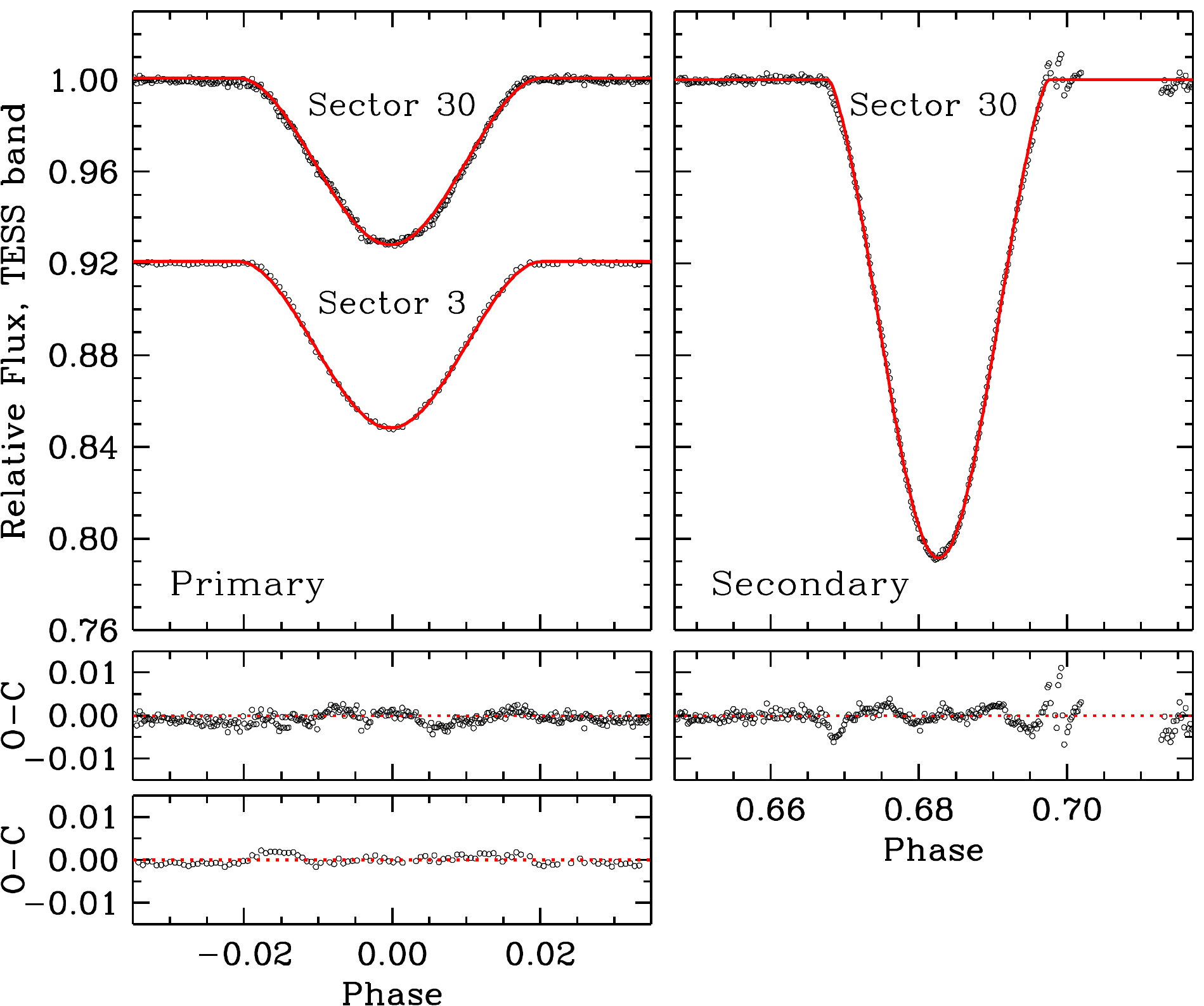}
    {\vskip 10pt
    \includegraphics[width=0.95\linewidth,trim= 0 0 65 335,clip]{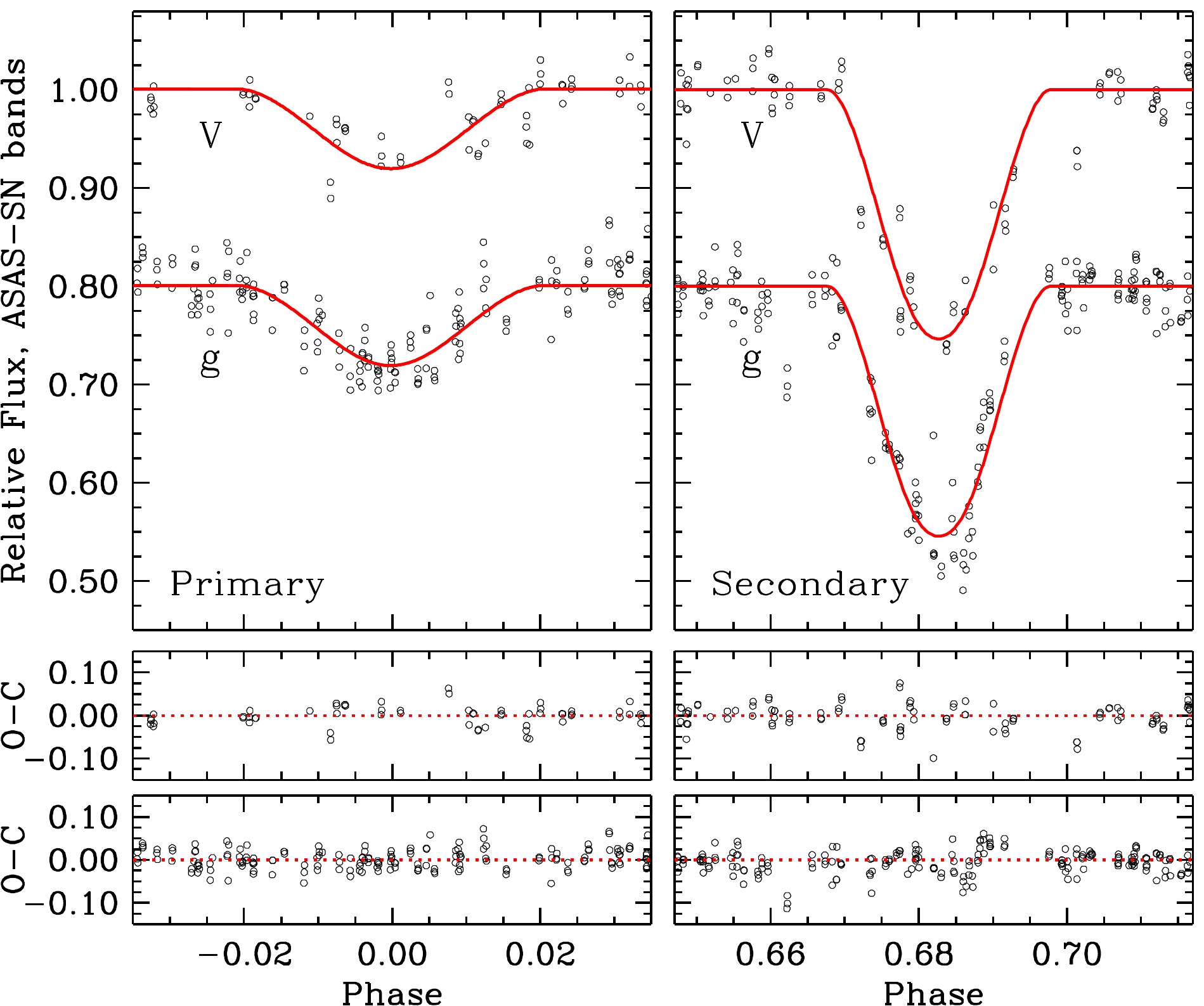}}
    \caption{Eclipse model shown with the TESS photometry for sectors~3
    and 30 (top) and the ASAS-SN $V$- and $g$-band measurements (bottom).
    Sector~3 and the $g$ band are displaced vertically for clarity. Residuals
    are shown below each panel. {They display systematic deviations that may be due to a combination of spots on one or both stars, stray light affecting the TESS photometry, and possible errors in our detrending.}}
    \label{fig:TESS_ASAS_models}
\end{figure}

A graphical representation of our solution is shown in the
top panel of Figure~\ref{fig:TESS_ASAS_models}, for the TESS photometry,
and in the bottom panel for the ASAS-SN $V$- and $g$-band observations. {The residuals within the eclipses display small systematic deviations from zero that are likely due in part to the spottedness of one or both stars. Other possible causes include subtle errors in the detrending, and stray light affecting one of the TESS eclipses but not the other (see Figure~\ref{fig:lc}). One consequence of such ``red noise", not accounted for in our MCMC analysis, is that our uncertainties for the fitted and derived properties {in Tables~\ref{tab:mcmc} and \ref{tab:derived}} may be underestimated to some extent.}

{As noted in Section~\ref{sec:fluxratios}, our spectroscopic flux ratios could be affected by the spottedness of the stars to a degree that is difficult to quantify,} so the uncertainties may be too small. This can propagate to other derived properties in this work. As a test, we loosened the flux ratio priors by arbitrarily multiplying the $\ell_2/\ell_1$ errors by a factor of three, and repeated the analysis. None of the fitted or derived quantities themselves changed significantly, and their uncertainties from our MCMC analysis increased by no more than a factor of {2}. Most of the uncertainties, such as those for the third light parameters and the radius sum, only increased by about {10--30}\%, and the errors on the geometric and dynamical properties ($P$, $T_0$, $e$, $i$) as well as those that depend more directly on the spectroscopy, such as the masses, {hardly changed} at all. We conclude from this experiment that even though our flux ratio errors may be underestimated, typically only about half of that bias transfers over to the uncertainties in the fitted and derived properties.

We note that the third light parameter for TESS is about {17}\%, whereas
the values for the ASAS-SN lightcurves are considerably smaller
($\ell_3 \approx {2}$--{4}\%) and statistically less significant. In
principle this difference could be explained if there is extra flux
from nearby stars in the much larger aperture of TESS compared to
ASAS-SN, but in \eb\ we do not believe that is the case. Contamination
by nearby stars was explicitly accounted for in the extraction of the
TESS lightcurves (see Section~\ref{subsec:LC}). Instead, we attribute
the third light to the fact that part of the
sector~30 photometry was affected by stray light from the Earth and/or
the Moon (cadences with quality flag 2048), biasing the background
flux level. As mentioned earlier, unfortunately this problem happened
precisely during the secondary (deeper) eclipse in sector~30, beginning about
halfway down the descending branch and lasting for the remainder of
that event (see Figure~\ref{fig:lc}). As this is the only secondary
eclipse present in the TESS data, we were left with little choice but
to retain the affected data for our analysis. The third light value
for TESS is therefore not representative of true stellar flux from
neighboring stars. The $\ell_3$ values for the $V$ and $g$ bands, on
the other hand, possibly do correspond to extra light from one or more
unseen companions in their much smaller apertures \citep[which are typically
about 16\arcsec\ in radius; see, e.g.,][]{Kochanek:2017}, consistent with the $\sim$1.4\% contamination estimated from the TIC (see Section~\ref{subsec:LC}), though we note again that formally the best-fit $\ell_3$ values are barely significant at the $\sim$1$\sigma$ level.

\begin{figure}[!t]
    \centering
    \includegraphics[width=\linewidth,trim=8 8 0 8,clip]{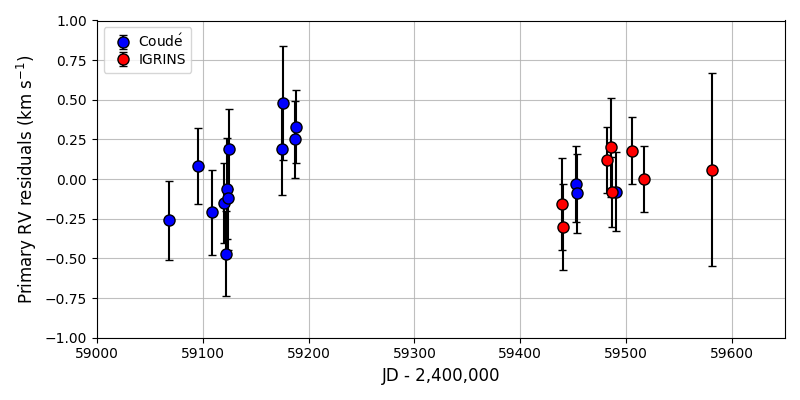}
    \caption{RV residuals from our model described by the parameters
    in Table~\ref{tab:mcmc}, after subtracting the motion of the EB. Only the
    primary velocities from the Coud\'e and IGRINS instruments are displayed for clarity,
    as the secondary residuals have a large scatter on the scale of this figure.}
    \label{fig:rvresid}
\end{figure}

Preliminary spectroscopic-only orbital solutions for \eb\ showed a
hint of a systematic trend in the residuals on a timescale of hundreds of days, particularly for the more
precise primary RVs from the Coud\'e spectrograph (see Figure~\ref{fig:rvresid}). This may suggest
the presence of a third body in the system. This pattern persisted in
the solution presented above, prompting us to explore it in more
detail. However, our attempts to add free parameters to model the
motion of the eclipsing binary in the outer orbit yielded results with
low statistical significance, likely because of the relatively small
number of observations and poor phase coverage.

In light of the significant spot-related activity in the system, including photometric variability that also appears on a timescale of hundreds of days (see Section~\ref{sec:continuum}), we consider it possible that the RV residuals are a manifestation of magnetic activity.

\subsection{Long-term photometric variability}\label{sec:continuum}

\begin{figure}[!t]
    \centering
    \includegraphics[width=\linewidth]{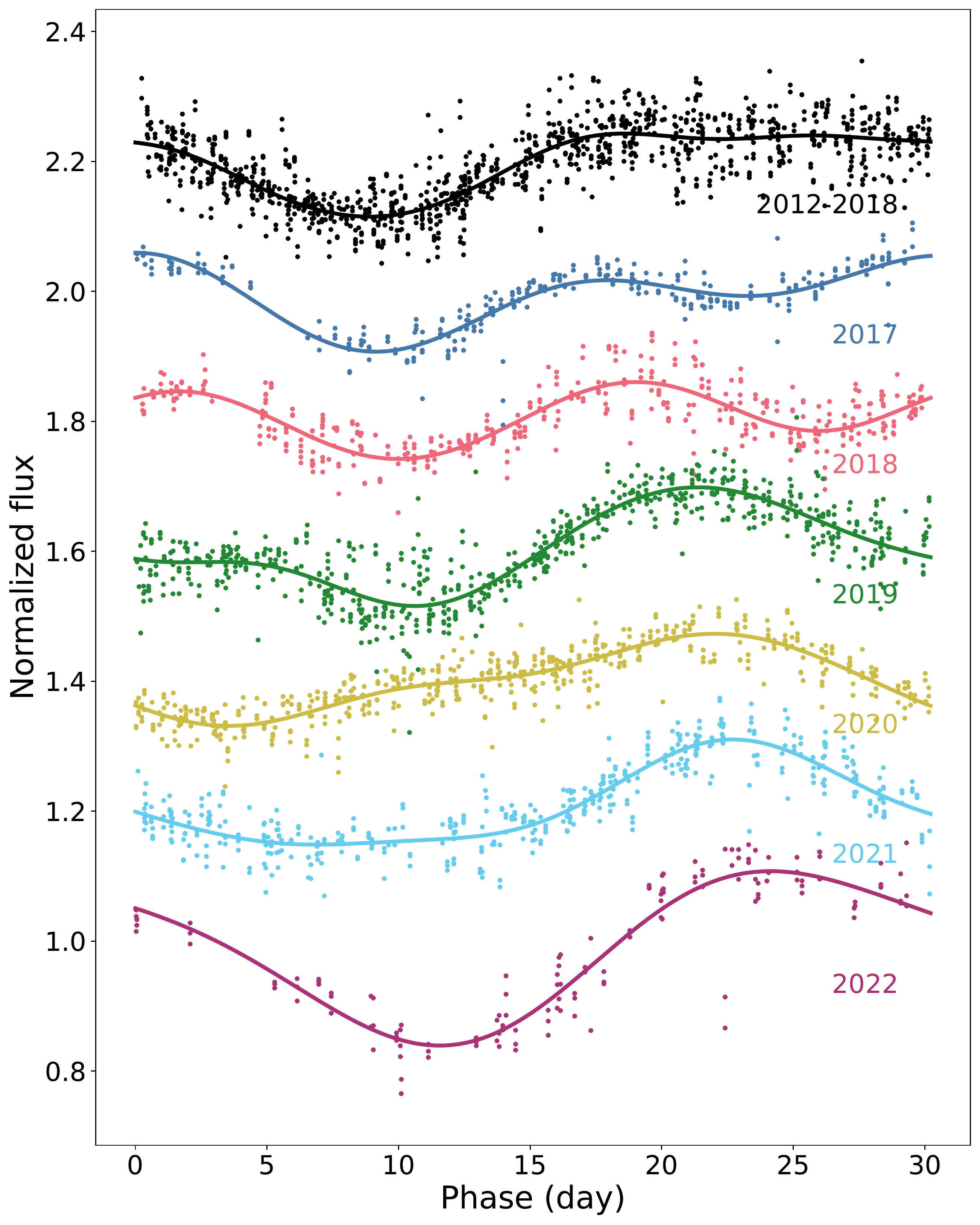}
    \caption{Phase folded ASAS-SN light curves, arbitrarily offset from one another, showing the evolution of the spot-induced variability over time. The black line shows the relatively stable V-band light curve, others show the g band light curve split by the observing season.
}
    \label{fig:spotevolution}
\end{figure}

\begin{figure}[!t]
    \centering
    \includegraphics[width=\linewidth,trim=10 5 50 38,clip]{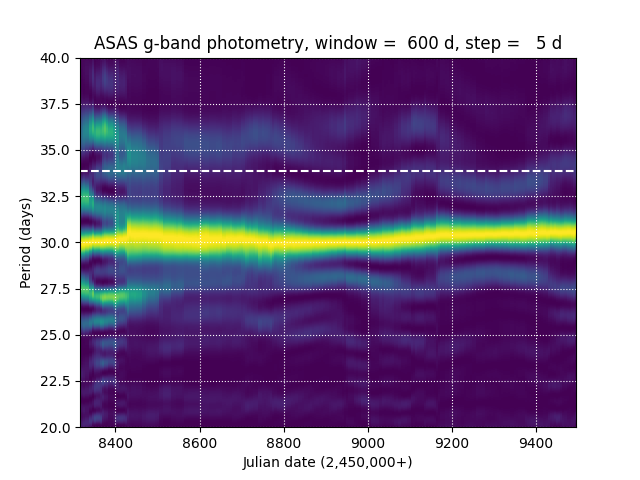}
    \includegraphics[width=\linewidth,trim=10 5 50 38,clip]{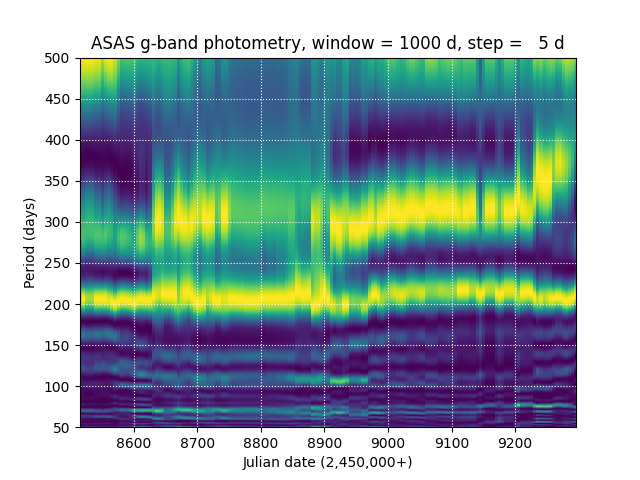}
    \caption{Sliding Lomb-Scargle periodograms of the ASAS-SN $g$-band data for periods in the range 20--40 days (top) and 50--500 days (bottom). The periodicity at 30.2 days (top) corresponds to the rotation period of the eclipsing stars (see Figure~\ref{fig:spotevolution}), which is shorter than the orbital period of 33.9 days (white dashed line). The periodicity at $\sim$210 days (bottom) corresponds to the long-term trend seen in the photometric variability after removal of the rotation period signal (see Figure~\ref{fig:longtend}).}
    \label{fig:lomb}
\end{figure}

In addition to the eclipses, \eb\ shows variability from spots on the surface of one or more of the stars. 
Given the long timescale of the variability, it is best characterized in the long-timescale ASAS-SN data, where it is seen throughout all 10 years of available data. From the coherent periodic signal in these data (Figure~\ref{fig:spotevolution}), we infer the rotational period of the star(s) responsible to be 30.22 days (Figure~\ref{fig:lomb}, top), comparable to but approximately 11\% shorter than the EB orbital period. 

There is little doubt that this 30.2~day period represents the rotation period of both stars. Using the spectroscopically measured \vsini\ of both stars (Section~\ref{subsec:vsini}) together with the precisely measured stellar radii (Section~\ref{subsec:lcfit}) yields the stars' rotation periods to be {30.6}$\pm${0.3}~d and {30.2}$\pm${0.6}~d for the primary and secondary, respectively, identical to $\sim$1\%. We defer to Section~\ref{sec:discussion} a discussion of the remarkable fact that the stars rotate synchronously.

\begin{figure*}[!ht]
    \centering
    \includegraphics[width=\linewidth]{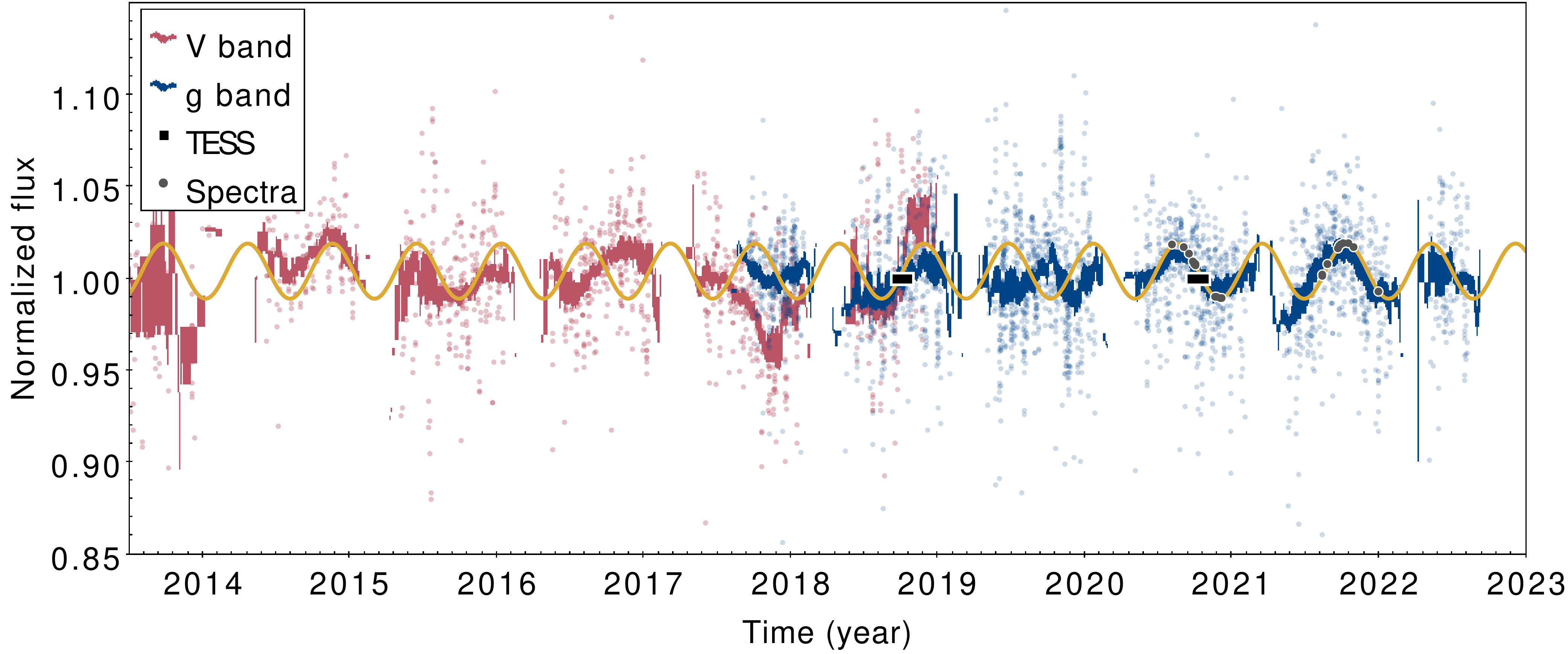}
    \caption{Long-term trends in the variability of \eb, excluding the eclipses, and removing the rotation signatures from Figure \ref{fig:spotevolution}. The thick line shows the smoothed median of the data in each band. The circles show the times of the spectroscopic observations. The yellow line shows the sinusoid with the period of $\sim$210 days approximating the light curve.
}
    \label{fig:longtend}
\end{figure*}

While this rotational signal presents the same period over the 10 years of the ASAS-SN data, there is a significant evolution in the morphology of the light curve (Figure~\ref{fig:spotevolution}). This evolution is not strongly apparent in the ASAS-SN $V$-band data, which was obtained from 2012 through 2018. However, with the observations conducted in the $g$ band, there was a continuous change, with the initial morphology of the light curve resembling that of the $V$-band data, with a single pronounced dip, eventually developing a secondary dip. The primary dip then transformed into a plateau, and currently the light curve has transitioned to simpler sinusoidal variation. This is almost certainly due to the evolution of the location or morphology of spots on the photosphere(s). 
The $g$-band and $V$-band data mostly sample different 5-year spans (Figure~\ref{fig:longtend}) and therefore the differences in morphological changes between them may reflect changes in spot evolution over the full 10-year timespan of the combined dataset. Some differences in behavior are observed during the brief period of overlap, which may potentially be due to differences in the dominant sources of opacity in the narrow $g$-band versus the wide $V$-band.

We fit this rotational variability with a 20th-order polynomial in the phase-folded light curve in each of the ranges shown in Figure~\ref{fig:spotevolution}. Subtracting the fitted function reveals a longer term variability in the system as well, with an amplitude of a few percent of the total flux (Figure~\ref{fig:longtend}). These fluctuations are quasi-periodic with a timescale on the order of $\sim$210 days (Figure~\ref{fig:lomb}, bottom). The precise timescale is difficult to determine, in part due to a lack of strong periodicity, and in part due to the gaps in the data when the system was inaccessible to a ground-based instrument. 

\subsection{Spectral Energy Distribution}\label{subsec:sed2}

With the benefit of the joint solution to the light-curve and radial-velocity data, we redid the SED fit from Section~\ref{subsec:sed1} now using two stellar photospheres, their individual \teff\ and $R$ informed by the eclipse modeling and the total system flux enforced to be consistent with the tight {\it Gaia\/} distance constraint (see Section~\ref{sec:eb}), 
and then also including a small contribution of light from a faint third source, detected in the form of 
excess emission in the (activity-corrected) FUV and NUV photometry. 
The resulting \teff\ and $R$ were iteratively updated based on the joint light-curve and radial-velocity model (Section~\ref{subsec:lcfit}) until a final satisfactory SED fit was produced. 
{To account for potential unmodeled sources of systematic error, the uncertainties on the broadband flux measurements were inflated in the usual manner to enforce reduced $\chi^2$ of unity for the final best fit; the resulting inflation factor of 3.2 is tantamount to increasing the reported photometric uncertainties from typically 0.01--0.02 mag to $\sim$0.05 mag. This also has the effect of increasing the formal uncertainties on the best-fit parameters from the SED fit.}

\subsubsection{Refined Effective Temperatures}

The relative eclipse depths measured from the light curves provide strong constraints on the ratio of stellar surface brightnesses ($J$) in the TESS, $V$, and $g$ bands 
(see Section~\ref{subsec:lcfit} and Table~\ref{tab:derived}), which in turn provide strong constraints on the ratio of \teff\ from the PHOENIX model atmospheres \citep{husser2013}. Note in particular that these surface brightness ratios require the nominal ``secondary" star to be the hotter component.

An additional constraint is provided by the eclipse durations, which very tightly constrain the sum of the radii of the eclipsing bodies, 
$R_1 + R_2$ ($R_{\rm sum}$)
(see Section~\ref{subsec:lcfit}). Another constraint is provided by the ratio of fluxes of the two stars ($\ell_2 / \ell_1$) from their relative broadening function peak areas in the spectra used to measure the radial velocities (Section~\ref{subsec:rv}), 
as summarized in Table~\ref{tab:derived}. Finally, the combined bolometric luminosities of the two stars via the Stefan-Boltzmann relation must reproduce the observed combined-light bolometric flux at Earth given the precise distance provided by {\it Gaia}. 


The resulting best-fit SEDs for the two eclipsing stars are represented in Figure~\ref{fig:sed} by the red and cyan curves, respectively, with best-fit $T_{\rm eff}$ of {4138}$\pm${52}~K and {4273}$\pm${38}~K, respectively, and best-fit $R$ of {4.671}$\pm${0.050}~\rsun\ and {3.932}$\pm${0.040}~\rsun.
These parameters successfully reproduce the $R_{\rm sum}$ constraint ({8.593}$\pm${0.059}~\rsun\ versus {8.592}$\pm${0.060}~\rsun\ from the final eclipse model; see Section~\ref{subsec:lcfit}), the flux ratio constraints ({0.847}$\pm${0.019} versus {0.847}$\pm${0.017}, {0.916}$\pm${0.022} versus {0.920}$\pm${0.024}, and {0.879}$\pm${0.022} versus {0.879}$\pm${0.021} in the TESS, $V$, and $g$ bands, respectively), the surface brightness ratio constraints ({1.184}$\pm${0.026} versus {1.190}$\pm${0.031}, {1.284}$\pm${0.038} versus {1.295}$\pm${0.047}, and {1.247}$\pm${0.024} versus {1.232}$\pm${0.038} in the TESS, $V$, and $g$ bands, respectively), and the {\it Gaia\/} distance ({677.2}$\pm${9.8}~pc versus 677.3$\pm$9.0~pc).

{Finally, we consider the effect of spots on one or both of the stars as a source of additional systematic uncertainty on \teff. Whereas \citet{Miller:2020} find a fundamental uncertainty floor of 11~K for the methodology used here when applied to quiet stars (due to irreducible uncertainties in the absolute flux scales of various photometric systems), the stars in \eb\ evince spot-induced photometric variations that are 8--9\% of the total system flux (Figure~\ref{fig:spotevolution}). If we assume conservatively that the variations arise entirely from one star accounting for $\sim$50\% of the total system flux, then we would infer intrinsic flux variations of about 16\% due to spots. This would correspond to an average variation in \teff\ of $\sim$80~K for a $\sim$4200~K star. Therefore, we add 80~K uncertainty in quadrature to the formal uncertainties determined above and adopt final \teff\ of 4138$\pm$95~K and 4273$\pm$89~K.}

\subsubsection{UV Excess: Tertiary Companion?}\label{subsec:tertiary}


As noted in Section~\ref{subsec:activity}, the \eb\ system is an X-ray source and also exhibits excess emission in the {\it GALEX\/} FUV and NUV bands, which can be largely understood as coronal and chromospheric in origin, consistent with the observed rotational properties of the stars. However, after accounting for the chromospheric contribution to the UV emission, a significant FUV excess remains (Figure~\ref{fig:sed}), which may be attributable to a hot, tertiary companion. 

As an initial estimate of the putative companion's properties, we fit a simple blackbody to the (activity-corrected) FUV and NUV fluxes (green curve in Figure~\ref{fig:sed}), assuming it is at the same distance as the \eb\ system and seen through the same extinction. 
The resulting best-fit parameters are $T \approx 11\,000$~K and $R \approx 0.03$~\rsun, which would be suggestive of an extremely low-mass white dwarf ($\approx$0.1~\msun). 
We revisit this possibility in Section~\ref{subsec:wd}.

\section{Discussion}\label{sec:discussion}

\subsection{Comparison to standard stellar models}\label{subsec:model_comp}

\begin{figure}[!t]
    \centering
    \includegraphics[width=\linewidth,trim=0 0 110 325,clip]{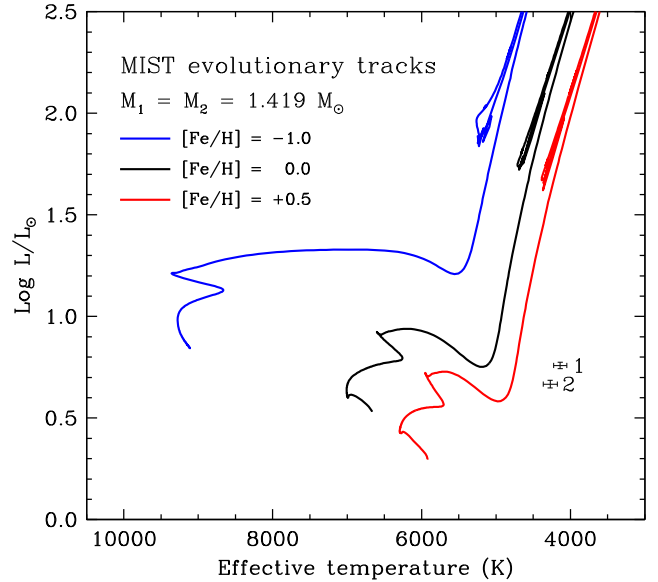}
    \caption{Comparison of the radii and temperatures of the eclipsing components in the \eb\ system against a MIST theoretical evolutionary track for the nearly identical masses of the components. As expected for RSSs, the stars appear far displaced to cooler \teff\ compared to normal subgiant or red giant stars of the same mass and metallicity.}
    \label{fig:tracks_comp}
\end{figure}

Consistent with our identification of \eb\ as a member of the unusual class of stars known as red stragglers, the measured properties of the eclipsing stars in the system are highly discrepant with those predicted by standard stellar evolutionary models. For example, Figure~\ref{fig:tracks_comp} compares the measured radii and \teff\ of the stars with the MIST evolutionary track for the measured mass of {1.419}~\msun\ (both stars have the same mass to within {0.2}\%). In particular, and consistent with the defining characteristics of the RSS class, the stars appear significantly cooler than expectation and/or may be significantly under-luminous. Note that even a super-solar metallicity (the system in fact has a somewhat subsolar metallicity of $-0.5 \pm 0.5$; Section~\ref{subsec:metallicity}) cannot resolve such a large discrepancy. 

\begin{figure}[!t]
    \centering
    \includegraphics[width=\linewidth,trim=0 0 105 175,clip]{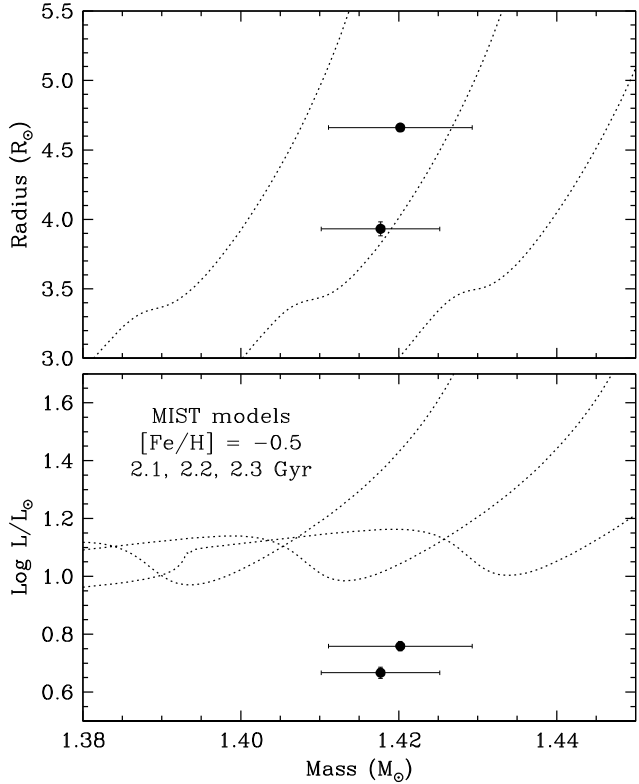}
    \caption{Same as Figure~\ref{fig:tracks_comp}, but comparing radius versus mass (top) and luminosity versus mass (bottom), for a range of ages consistent with the measured radii.}
    \label{fig:tracks_comp_mr}
\end{figure}

It is interesting that, despite these very large discrepancies relative to standard theory, the stars nonetheless appear qualitatively to follow the theoretical evolutionary track, as if simply displaced to cooler \teff\ and/or to lower luminosity. Indeed, comparing the stars to the standard models in mass-radius space (Figure~\ref{fig:tracks_comp_mr}, top) shows that one can find an age at which the stars indeed appear to follow a ``normal" mass-radius relationship (at an age of $\sim${2.2~Gyr} in the case of the MIST models); at this stage of evolution, the observed difference of $\sim$15\% in the stars' radii is in fact consistent with the very slight formal difference of {0.16}\% in the stars' masses. However, relative to the same models, the stars do indeed appear to be under-luminous by $\sim${0.4}~dex. In other words, we may regard the stars as obeying a normal mass-radius relationship, but their \teff\ are significantly suppressed and therefore they are also significantly less luminous than expected. 

It is an especially striking feature of the \eb\ system that the masses of the eclipsing stars are so remarkably similar, formally differing by only {0.18}\% and at most differing by $\sim${0.5}\% at 95\% confidence (Table~\ref{tab:derived}). To what extent are such identical masses among spectroscopic binaries to be expected? Studies of ``twins" among spectroscopic binaries have recently found evidence for a significant excess of systems with $q > 0.95$, suggesting that the most ``identical twins" may be formed via pathways that differ from binaries with less identical mass ratios. Estimates from these studies suggest that the occurrence of twins with $q > 0.95$ among spectroscopic binaries may be as high as $\sim$25\% \citep[see, e.g.,][]{El-Badry:2019}; for $q > 0.98$ it is estimated at $\sim$3\% \citep[see, e.g.,][]{Simon:2009}. Thus, we may conservatively infer that the occurrence of truly identical twins like \eb, with $q >$ {0.998}, is much less than a few percent.

\begin{figure}[!t]
    \centering
    \includegraphics[width=\linewidth,trim=0 0 95 340,clip]{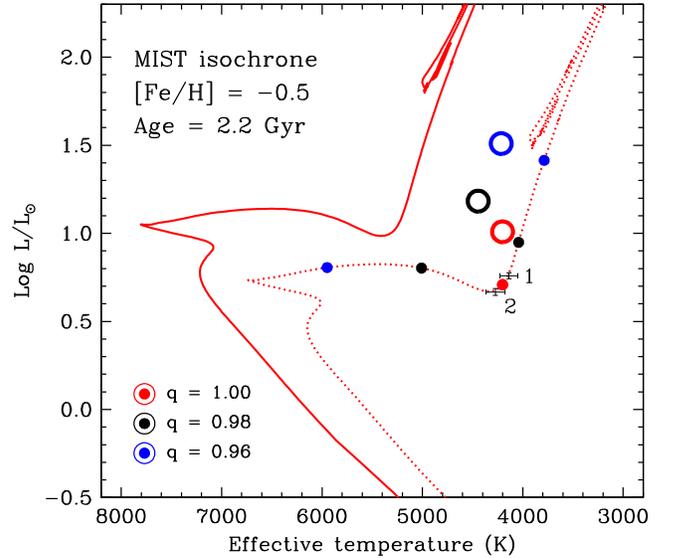}
    \caption{The components of simulated binaries in the HR diagram, selected to have the same total mass and orbital period as \eb, but for different choices of the mass ratio, $q$, and requiring the stars' radii to remain detached. Each pair of component stars is represented by filled symbols of the same color on a standard stellar evolution model, which has been manually shifted in \teff\ and $L$ to crudely simulate the suppressed temperatures and luminosities observed (dotted curve); the resulting combined-light position for that pair is represented by the open symbol of the same color. In all cases, the combined light of the synthetic binaries falls in the RSS region of the diagram (i.e., to the right of the standard model track; solid curve). For the \eb\ system to have been selected as an RSS EB candidate implicitly required $q \gtrsim 0.95$; see the text.}
    \label{fig:tracks_comp_bin}
\end{figure}

To what extent might the identical masses of \eb\ represent an observational bias due to the selection function for the RSS class? The \eb\ system was not selected randomly for our investigation; it was selected because it appears in the RSS region of the HR diagram (Section~\ref{sec:eb}), and this condition may imprint a bias on the range of $q$ that an EB in that region can possess. 
Figure~\ref{fig:tracks_comp_bin} attempts to assess this by simulating the HR diagram positions of binaries with the same total mass and orbital period as \eb, but with different $q$ and requiring the stars' radii to remain detached (we draw the component stars from a standard stellar evolutionary track, manually shifted in \teff\ and $L$ to crudely simulate the observed suppressed \teff\ and luminosities for this exercise). 
We find that for an EB like \eb, if $q \lesssim 0.95$, the larger star would reach the tip of the red giant branch and fill its Roche lobe, and thus would no longer appear as a well detached EB. For $q \gtrsim 0.95$, the stars remain detached and the system remains in the general vicinity of the RSS region in the HR diagram. 

Thus, we may conclude that the selection of \eb\ as a candidate RSS EB may have precluded system configurations with $q < 0.95$. However, this would have still permitted the \eb\ system in principle to have any mass ratio within the range commonly observed for ``twins" \citep[$\sim$25\% occurrence for $q > 0.95$;][]{El-Badry:2019} or even that observed for ``identical twins" \citep[$\sim$3\% occurrence for $q > 0.98$;][]{Simon:2009}. Our finding of extremely similar masses ($q >$ {0.998}) for the \eb\ system does not appear to have been the result of an observational bias or selection effect.

A number of other well-studied EBs with such identical masses are known. One example involving evolved stars with a complex history is the double blue-straggler system reported by \citet[][star identifier 7782]{Mathieu:2009} with $q = 1.005 \pm 0.013$. In addition, at least two identical twin EBs are known at pre--main-sequence ages, implying that identical twin binaries can emerge during or very soon after the star-formation process. For example, \citet{Tofflemire:2022} report a low-mass EB with $q = 1.000 \pm 0.001$ at an age of $\sim$40~Myr, and \citet{Stassun:2008} report a low-mass EB with $q = 0.99 \pm 0.02$ at an age of only 1--2~Myr (i.e., almost immediately after the end of the protoplanetary disk accretion phase). The latter system was also later found to possess a tertiary companion that may have played a role in that system's properties and evolution \citep[see][]{Gomez:2012}.

\subsection{Alternative Evolutionary Models}
\begin{figure*}[!t]
    \centering
    \includegraphics[width=\linewidth,trim=700 500 700 500,clip]{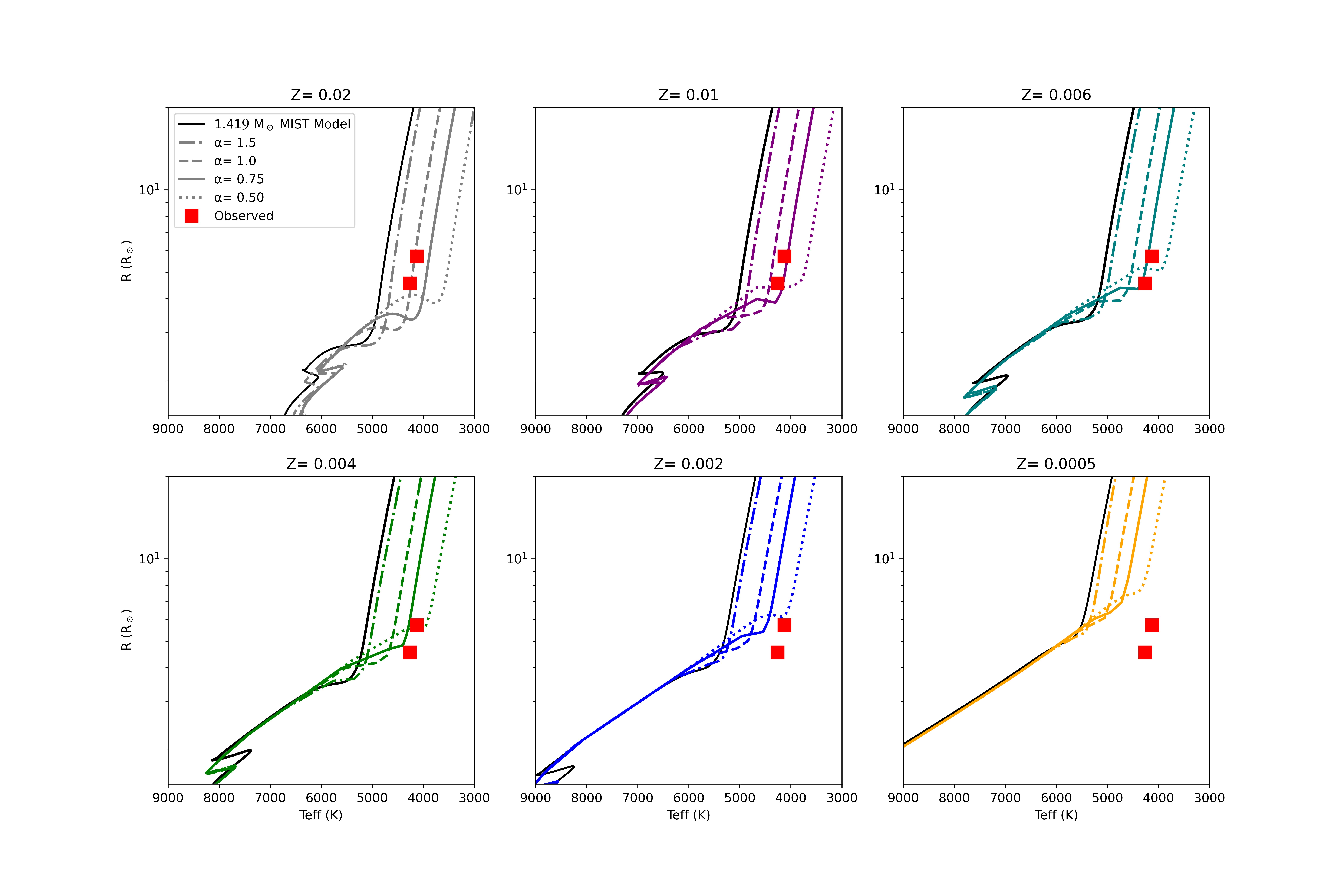}
    \caption{Alternative evolutionary models incorporating the effects of reduced convective efficiency due to a large starspot covering fraction. The solid black track shows standard MIST evolutionary tracks for a {1.419}~\msun\ star, which use a standard solar calibrated mixing length $\alpha = 1.82$ \citep{Choi:2016}. Colored tracks are MESA evolutionary tracks of varying metallicity that use adjusted $\alpha$ values as indicated.}
    \label{fig:tracks_comp_alt}
\end{figure*}

{The observed properties of the eclipsing stars in \eb\ may be understood as the result of both
(i) interior effects on convective efficiency due to strong internal magnetic fields and (ii) surface effects due to large starspot covering fractions.} In magnetically active stars, convective motions of fluid across magnetic field lines will be preferentially opposed, slowing convective motions and diverting this kinetic energy into the magnetic field. One approach to model this effect on stellar structure is to reduced the convective mixing length parameter, $\alpha$ (\citealt{Chabrier:2007}; see also \citealt{Leiner:2017}). Figure~\ref{fig:tracks_comp_alt} shows MESA models for a range of metallicities and convective efficiencies, $\alpha$ (the convective mixing length parameter). For solar and slightly sub-solar metallicities ($Z = 0.02$, $Z = 0.01$, $Z = 0.006$), $\alpha$ between 1.0 and 0.7 yields an evolutionary track that nicely reproduces the observed \teff-radius relationship for \eb. For more metal poor models ($Z = 0.004$, $Z = 0.002$, $Z = 0.0005$), the reduced mixing length models cannot reproduce the observed radius and \teff\ of \eb. Standard metal-poor giants are hotter and larger than solar-metallicity giants, so if the observed binary is indeed quite metal poor, the observed discrepancy in temperature and radius is too large for the lowered mixing length models to reproduce. 

The ability of these models to reproduce the observed \teff\ and luminosities provides insight into the physical mechanisms by which activity acts to alter the global and interior properties of cool, evolved stars. 
EB-based studies of these effects in the pre--main-sequence (PMS) and main-sequence phases have similarly shown that rotationally driven activity produces decreased surface temperatures and inflated radii compared to standard, non-magnetic stellar models \citep[see, e.g.,][]{Stassun:2012}; these studies additionally find that ``magnetic" stellar models, which parameterize the effects of surface fields through reduced convective efficiency \citep[e.g.,][]{Feiden:2016}, are able to reproduce the suppressed temperatures and inflated radii \citep[see, e.g.,][]{Stassun:2022}. There is also good evidence that the interior temperature structure through the convective zone is altered, such that surface Li abundances in active PMS stars are far less depleted than expected from non-magnetic stellar models \citep[see, e.g.,][]{Stassun:2004,Stassun:2022}. 

However, in the case of the PMS and main-sequence stars, these alterations of the global and interiors properties of the stars appear to leave the stellar luminosity intact; that is, the surface temperature and radius essentially adjust to transport via reduced convective efficiency the same luminosity emerging from the core \citep[see, e.g.,][]{Stassun:2012,Stassun:2022}.


Apparently, in the case of evolved stars like the red giants in \eb, the effect of such activity may be more profound, altering (reducing) the stellar luminosity and resulting in a reduction of the surface temperature (but leaving the stellar radius largely unaffected; see Figure~\ref{fig:tracks_comp_mr}). 
\citet{Leiner:2017} suggest, based on their exploration of models with reduced convective mixing lengths,
that lowered mixing lengths create cooler, larger stars at all points during the evolution of the star, with the luminosity remaining unchanged at most stages of
evolution regardless of mixing length. However, altered mixing lengths do create lower-luminosity stars near the end of the subgiant branch and the beginning of the red giant branch. At this stage in evolution, the expanding shell absorbs enough energy to lower the luminosity for a time. Lowering the mixing length parameter further leads to an even greater dip in the luminosity on the red giant branch, and this luminosity dip occurs at lower \teff. In the specific case of the \eb\ system, the luminosity dip is best matched by $\alpha$ between {0.5}--1.0, depending on the adopted metallicity (see Figure~\ref{fig:tracks_comp_alt}).

It is interesting that these effects during the pre--main-sequence (PMS) phase are more similar to the main sequence than to the subgiant and red giant phases.
Perhaps this is because PMS stars and red giants do not liberate energy in quite the same way. Whereas PMS stars undergo global contraction, red giants undergo more complex changes, wherein  at this particular phase of evolution the star is thermally readjusting after the end of hydrogen burning and some of the luminosity is being deposited in the expanding envelope \citep[see, e.g.,][]{Leiner:2017}. 


More generally, these observed differences in the way that magnetic reduction of convective efficiency manifests across different stages of stellar evolution may help inform how future stellar models incorporate magnetic effects on global stellar properties and on the physics of stellar interiors \citep[see also, e.g.,][]{Stassun:2004,Stassun:2012,Stassun:2022}.

\subsection{Interpretation of System Properties and Evolutionary History}

\subsubsection{Rotational Evolution}
In addition to their strikingly similar masses, another remarkable feature of the \eb\ system is the eclipsing stars' very similar rotation periods ($P_{\rm rot} \approx 30$~d for both stars; see Section~\ref{sec:continuum}). It is well established that stars in close binaries will tidally interact and consequently synchronize their rotation. In the case of non-circular orbits, the stars come to rotate ``pseudo-synchronously" {\citep[e.g.,][]{Hut:1981}}.
For \eb, the pseudo-synchronous period is $\sim$19~d {\citep[see Eq.~42 in][]{Hut:1981}}, about 30\% faster than the stars currently rotate. 

Studies of the rotational (pseudo-)synchronization of binaries suggest that the stars must be close to one another, and that the most relevant measure of ``closeness" is the stars' radii relative to the orbital separation (i.e., the stars' relative radii, $r$). In their detailed review of the properties of nearly 100 benchmark EBs spanning a range of evolutionary states, \citet[][c.f.\ their Figure~10]{Torres:2010} showed that the stars achieve pseudo-synchronous rotation when $r \gtrsim 0.09$ and that, for cool stars with convective envelopes, they experience spin-up toward pseudo-synchronous rotation when $0.04 \lesssim r \lesssim 0.09$. Tidal effects appear to be ineffective for well-separated stars with $r \lesssim 0.04$. 

The eclipsing stars in \eb\ currently have relative radii of $\approx$0.07 (see Table~\ref{tab:derived}), thus suggesting that they should be experiencing tidal spin-up toward pseudo-synchronous rotation. Assuming they continue to evolve up the red giant branch, they will reach $r = 0.09$ when they are $\sim$25\% larger, within the next $\sim$50~Myr (see Figure~\ref{fig:tracks_comp_mr}, top). 

Observations and models of the rotational evolution of stars through the subgiant and early red giant stages of evolution clearly show that, by the time normal stars reach the base of the red giant branch, they should rotate very slowly. For example, CoRoT observations of evolved stars by \citet{doNascimento:2012} find typical rotation periods at the base of the red giant branch in the range 50--100~d; for stars with masses $\sim$1.5~\msun\ similar to the stars in \eb, their observations suggest a most likely range of $90 \pm 10$~d. Thus, the rotation period of 30.2~d we observe in \eb\ is a factor of $\sim$3 faster than would be expected for single (or well-separated binary) stars near the base of the red giant branch. According to \citet{doNascimento:2012}, the rotation period we observe in \eb\ would normally occur between the time a star turns off the main sequence and traverses the ``blue hook". Interestingly, assuming the orbital separation of \eb\ has not changed significantly in the past, the eclipsing stars would have increased their radii to $r \gtrsim 0.04$ around that time. 

Therefore, one possibility is that the stars in \eb\ first began to interact tidally around the time that they would have individually spun down to a rotation period of $\approx$30~d near the blue hook. Then, they could have maintained a more rapid rotation than would otherwise be expected, as their relative radii have increased and their tidal interaction continues to increase further. Very soon, as noted above, the stars will reach the $r \gtrsim 0.09$ threshold and presumably by then will have fully achieved spin-up to the pseudo-synchronous rotation period of $\sim$19~d. In this scenario, then, the eclipsing stars in \eb\ are presently in a state of significant and increasing tidal interaction.

This picture begs the question, however, of why the stars already share a common rotation period, i.e., why they are already rotationally synchronized.
On the main sequence, these stars with $M \approx 1.42$~\msun\ would have been well separated with $r \approx 0.02$ (again assuming the same orbital separation as at present), so a common rotation period due to tidal interaction would not have been expected at that stage. 
%
In addition, because the stars in \eb\ would have been early F- or A-type stars (depending on the metallicity) when they were on the main sequence, they would not be expected to participate in the well-behaved mass-rotation relationship of cooler stars. 

Perhaps the answer is as simple as, with the stars now tidally interacting and proceeding to pseudosynchronization, tides were able to synchronize the stars with one another along the way.
However, most binaries with similar orbital periods 
as \eb\ are not observed to be so rotationally synchronized. 
For example, \citet{lurie2017} examined some 800 EBs observed by {\it Kepler\/} showing rotationally modulated spot variations, 13 of which have rotation periods measured for both stars in the EB and orbital periods in the range 20--50~d, comparable to the 33~d orbital period of \eb. None of these are synchronized to within 1\% as in \eb\ (see Section~\ref{sec:continuum}). However two of them are synchronized to within 3\% and 6 of them to within 10\%, and \citet{lurie2017} suggest that one or both stars in those EBs may be evolved. 
Thus, for the orbital period of \eb\
and the stars' state of evolution, 
rotational synchronization may occur as frequently as 15--50\%, depending on the tolerance permitted for the synchronization (3--10\%). 

Why some binaries like \eb\ attain synchronized rotation with one another even before achieving full pseudo-sychronization with the orbit, while most binaries do not, is unclear. 
Perhaps 
the stars in \eb\ 
have shared very similar rotation periods from the beginning.
In that case, the stars' identical rotational properties, like their identical masses, would be interpreted as an outcome of the binary formation process (see also Section~\ref{subsec:model_comp}).

\subsubsection{On the Possibility of a White Dwarf Companion}\label{subsec:wd}


As mentioned in Section~\ref{subsec:model_comp}, the occurrence rate for spectroscopic binaries with a mass ratio of unity similar to \eb\ is expected to be well below 3\%. Recently, \citet[][PZL]{Portegies:2019} put forward a hypothesis for mass transfer in triple systems that they suggest may yield such equal companion masses. 

In this scenario, an evolved wide tertiary companion fills its Roche lobe, and the resulting mass transfer forms a circumbinary accretion disk around a closer inner binary. 
The lower-mass star in the inner binary, by virtue of being closer to the inner edge of the disk, will experience a higher accretion rate than the higher-mass star, driving the mass ratio of the inner binary toward unity. In addition to the resulting equal-mass inner binary, the PZL mechanism would predict the tertiary leaving behind a white dwarf remnant in a wide, coplanar orbit. 
By construction, PZL assume the tertiary is coplanar for reasons of dynamical stability.

The SED of the \eb\ system (Figure~\ref{fig:sed}) shows excess emission in the FUV which, as discussed in Section~\ref{subsec:tertiary}, may be explained by a hot tertiary companion in the system with $T \approx 11\,000$~K and $R \approx 0.03$~\rsun, which would be suggestive of an extremely low-mass (ELM) white dwarf ($\approx$ 0.1~\msun). 

One potential concern with attributing both the equal masses of the inner binary and the FUV excess to an ELM white dwarf is that, to produce such a low-mass white dwarf, the mass loss of the donor would have to happen early, such as during the early subgiant phase of the tertiary \citep{Li:2019}. This, in turn, would require a very close lower-mass companion to the donor star \citep{Sun:2018}, in other words an initially quadruple system. Whether the PZL mechanism would still operate on the inner binary is unknown. 

As noted in Section~\ref{subsec:sed2}, the observed FUV excess emission might not be from a hot companion but instead another manifestation of the apparently strong magnetic activity in the system. In that case, there would no longer be any direct evidence for a (low mass) white dwarf tertiary in the system. However, the presence of an undetected, normal-mass, cool white dwarf---as required by PZL---remains possible. Further investigation into the presence of a white dwarf tertiary is warranted.

\subsubsection{The \eb\ System in Context and the Nature of Red Straggler Stars}

{\eb\ is consistent with the observational definition of RSSs, as both component stars are redder than the expected RGB color for their derived masses (see Section~\ref{subsec:model_comp}). It is important to note that this definition is often conflated with that of SSGs, which are observationally defined to be redder than the main-sequence, but less luminous than the subgiant branch. \citet{Geller:2017b} distinguished between these two definitions by limiting the definition of RSSs to more luminous than the subgiant branch. By this definition, \eb\ could be either a SSG or RSS depending on the true metallicity, as seen in Figure~\ref{fig:tracks_comp}. The \citet{Geller:2017b} study also found that SSGs and RSSs both appear to be related to magnetically active evolved stars (RS CVn stars). For SSG systems with high cluster membership probability, they report 25/43 are X-ray sources and 28/43 are photometric or RV variable. For the RSSs, 4/7 were found to be X-ray sources and 3/7 had measurable photometric or RV variability. This supports the idea that systems in both areas of the HR diagram likely result from the evolution of magnetically active evolved binaries (RS CVn stars).}

{As stated in the Introduction, \citet{Leiner:2017} found that mass transfer, envelope stripping, and enhanced magnetic activity could all possibly create SSGs, but it also suggested that the magnetic activity channel was the most likely. This is further corroborated in the final paper of that series \citep{Geller:2017a}, in which enhanced magnetic activity had the largest analytic and Monte Carlo formation probability.}

{It remains unclear how common or unusual the \eb\ system is compared to other SSGs and RSSs. We are unaware of any other fully characterized SSG/RSS EBs, although current work to find them in the \citet{Leiner:2022} sample is in progress. We also note that we have not found results in the literature that might indicate a representative mass ratio distribution for this sort of binary system. In general, it appears that SB2s are not common among SSG/RSS binaries, as \citet{Geller:2017b} only report two of them, with mass ratios of 0.7 and 0.8 (the remainder of the sample being SB1s). Given the mass ratio of 1.0 in \eb, it may perhaps be argued that the \eb\ system is different from the typical SSG/RSS and that there must have been something special about its evolution.} 

{Regardless of its evolutionary history, the enhanced magnetic activity of \eb\ makes it suitable to serve as a benchmark system to test lowered mixing length models, as it is currently the only known SSG/RSS EB and the only known system of its kind with masses that are separately and empirically measured from the orbit.}

\section{Summary and Conclusions}\label{sec:summary}

We have discovered the \eb\ system as the first known eclipsing binary (EB) among the unusual classes of objects referred to as sub-subgiants (SSGs) and red straggler stars (RSSs). SSGs and RSSs are so named because of their large displacement in the HR diagram to cooler temperatures and lower luminosities relative to normal subgiants and red giants. 

Previous work has suggested that SSGs/RSSs may be products of close stellar encounters involving binaries, mass transfer, or stellar collisions \citep[e.g.,][]{Mathieu:2003,Albrow:2001,Hurley:2005}.
More recent studies \citep[e.g.,][]{Leiner:2017} have suggested that the reduced temperatures and luminosities may be due to magnetic fields that lower convective efficiency and produce large star spots, akin to the ``radius inflation" and ``temperature suppression" effects that have been reported in magnetically active young stars \citep[e.g.,][]{Stassun:2012,Tofflemire:2022} and in active, low-mass main-sequence stars \citep[e.g.,][]{Stassun:2022}. 
With directly and precisely determined masses, radii, temperatures, and luminosities, 
the \eb\ system provides a rare opportunity to help resolve these questions by directly linking the stars' evolutionary properties to their fundamental physical properties. 

Our measurements reveal the \eb\ system to possess a number of unusual characteristics that are not straightforward to understand. In particular, the masses of the two eclipsing stars are remarkably identical---both have masses of $\approx${1.419}~\msun\ to within {0.2}\%---and they rotate synchronously with one another (and nearly synchronously with the orbital period), which may be surprising given their relatively wide, eccentric orbit ($P_{\rm orb} = 33.9$~d, $e = 0.35$). 

The stars are clearly magnetically active, which manifests as strong NUV excess emission and very hard X-ray emission, presumably arising in chromospheres due to their rotation being much faster than would be expected for normal, single red giants. Stellar evolutionary models that have been modified to account for strong surface fields, in the form of reduced convective efficiency and large starspot covering fractions, are able to reproduce the suppressed temperatures and reduced luminosities that we observe. The key questions that remain, then, are the origins of the identical masses and sychronized rotation. 

There is evidence for excess FUV emission as well as long-term modulations in the radial-velocity data; it is not clear whether these features are also attributable to the activity of the RSSs or if they reveal a tertiary companion. 
If the system possesses a white dwarf tertiary, then a mass-transfer scenario in a triple system could explain the remarkably identical masses. As yet we have not been able to make a self-consistent model from our observations for such a white dwarf tertiary. This is a key question that followup studies should specifically investigate. 

Alternatively, the stars could be presumed to have been formed as identical twins, and that they managed to become tidally synchronized with each other as they evolved toward the red giant branch. 
The prevalence of ``identical twin" binaries is not precisely known, however there are a small number of other well-studied EBs known to possess identical masses, and interestingly some of them have very young ages of 1--40~Myr \citep{Stassun:2008,Tofflemire:2022}, proving that such identical masses can indeed arise very soon after the stellar birth process. 
In addition, studies of the large population of EBs discovered by {\it Kepler} found that, for the orbital period of \eb\ and its state of evolution, rotational synchronization may occur as frequently as 15--50\%, depending on the tolerance permitted for the synchronization \citep{lurie2017}.

If the identical masses and synchronized rotation can indeed be attributed to birth, then all of the features of the system can be explained via activity effects, without requiring a more complex dynamical history.

\acknowledgments
{We thank the anonymous referee for helpful comments on the original manuscript.}
We gratefully acknowledge partial support from the Vanderbilt Initiative in Data-intensive Astrophysics (VIDA). 
This study made use of the IGRINS instrument on the Gemini South telescope.
The international Gemini Observatory is a program of NSF's OIR Lab, managed by the Association of Universities for Research in Astronomy (AURA) under a cooperative agreement with NSF on behalf of the Gemini partnership: NSF (United States), National Research Council (Canada), Agencia Nacional de Investigaci\'on y Desarrollo (Chile), Ministerio de Ciencia, Tecnolog\'ia e Innovaci\'on (Argentina), Minist\'erio da Ciência, Tecnologia, Inovações e Comunicações (Brazil), and Korea Astronomy and Space Science Institute (Republic of Korea).
All of the TESS data presented in this paper were obtained from the Mikulski Archive for Space Telescopes (MAST) at the Space Telescope Science Institute. The specific observations analyzed can be accessed via \dataset[10.17909/t9-nmc8-f686]{https://doi.org/10.17909/t9-nmc8-f686}, \dataset[10.17909/3y7c-wa45]{https://doi.org/10.17909/3y7c-wa45},  \dataset[10.17909/fwdt-2x66]{https://doi.org/10.17909/fwdt-2x66}.

\bibliography{final}

\end{document}